\documentclass[journal=jcisd8,manuscript=article]{achemso}

\usepackage[T1]{fontenc}
\usepackage[utf8]{inputenc}
\usepackage{lmodern}
\usepackage{amsmath}
\usepackage{graphicx}
\usepackage{booktabs}
\usepackage{multirow}
\usepackage{longtable}
\usepackage{float}
\usepackage{xcolor}
\usepackage[normalem]{ulem}

\graphicspath{{./figures06/}{./}}

\title{Low-Temperature Transport in Li-Ion Battery EC/EMC/FEC Electrolytes: Molecular Dynamics and Machine-Learning Modeling}

\author{\.{I}pek Yenda \c{C}\i nar}
\affiliation{Chemical Engineering Department, Gebze Technical University, Kocaeli, T\"{u}rkiye}

\author{Oguzhan Orhan}
\affiliation{Department of Physics, F\i rat University, Elaz\i\u{g}, T\"{u}rkiye}

\author{M. Olu\c{s} \"Ozbek}
\affiliation{Chemical Engineering Department, Gebze Technical University, Kocaeli, T\"{u}rkiye}

\author{\c{S}ener \"Oz\"onder}
\affiliation{Institute for Data Science \& Artificial Intelligence, Bo\u{g}azi\c{c}i University, \.{I}stanbul, T\"{u}rkiye}
\email{sener.ozonder@bogazici.edu.tr}

\begin{document}

\maketitle

\begin{abstract} 
Low-temperature operation imposes severe limitations on lithium-ion transport in battery electrolytes, yet the coupled effects of solvent composition and fluorinated additives in the cold-temperature regime remain insufficiently resolved. Here, we combine classical molecular dynamics (MD) and machine learning (ML) to investigate 1~M LiPF$_6$ electrolytes containing ethylene carbonate (EC), ethyl methyl carbonate (EMC), and EC/EMC (3:7), with 0-10~mol\% fluoroethylene carbonate (FEC), from 298 to 233~K. MD simulations quantify Li$^+$ self-diffusion, Nernst-Einstein (NE) and Green-Kubo (GK) conductivities and local coordination, while Gaussian-process surrogates model conductivity across composition and temperature. Cooling produces a pronounced transport penalty, particularly in EMC-containing electrolytes, whose GK conductivity decreases by more than 98\% at 233~K, compared with approximately 90\% in EC-rich systems. Li$^+$ self-diffusion activation energies are 0.49-0.54~eV for EMC-containing systems and 0.27-0.29~eV for EC-based systems. Within the EC family, 0-2~mol\% FEC gives comparable cold-temperature transport, whereas 5-10~mol\% FEC shows lower conductivity retention at the coldest simulated temperature. The analyzed coordination channels remain solvent dominated, while direct Li$^+$-FEC coordination is not quantified in the present RDF set.The Gaussian-process surrogates achieve composition-disjoint cross-validated RMSE values of 0.56 and 0.57~mS~cm$^{-1}$ for NE and GK conductivity. Within the simulated liquid-state trajectories, temperature and host-solvent composition dominate the bulk-transport response, with FEC acting as a secondary modifier. 
\end{abstract}

\section{Introduction}

Lithium-ion (Li-ion) batteries are used in applications ranging from portable electronics to electric vehicles, renewable-energy systems and aerospace platforms \cite{diouf2015,zubi2018}. Their technological relevance arises from their high energy density, long cycle life, low self-discharge rate and high charge-discharge efficiency \cite{lyu2021,zhang2024beyond}. Despite these advantages, Li-ion battery performance is highly sensitive to temperature. Under cold conditions, slower charge-transfer kinetics, increased electrolyte viscosity, higher interfacial resistance, and an increased propensity for Li plating collectively degrade capacity, power output, and cycle life, underscoring the critical role of electrolyte design in mitigating these performance losses \cite{zhu2015materials,Li2017,Li2024,Belgibayeva2023,Piao2022}.

Beyond serving as the transport medium, the electrolyte controls ion solvation, electrochemical stability and solid electrolyte interphase (SEI) formation \cite{Liu2022,Cheng2024,Sun2022,Tang2022}. LiPF$_6$ remains the most widely used lithium salt because it provides high ionic conductivity and acceptable compatibility with conventional carbonate solvents, although its thermal instability and sensitivity to moisture motivate continued interest in electrolyte optimization and alternative salts \cite{Liu2022,Mahmud2022}. Among carbonate solvents, ethylene carbonate (EC) is particularly important because its high dielectric constant favors Li$^+$ dissociation and stable SEI formation on graphite. Its high viscosity and relatively high melting point, however, are unfavorable for low-temperature transport \cite{Wang2023,Su2023,Liu2022,Tang2022,Mahmud2022}. Ethyl methyl carbonate (EMC), by contrast, has lower viscosity and a lower melting point, and is therefore commonly mixed with EC to reduce transport resistance at low temperature \cite{Ringsby2021,Zhang2002}. EC/EMC mixtures, particularly those with an EC:EMC mass ratio of 3:7, are widely used because they balance ion dissociation, SEI formation and bulk mobility \cite{Ringsby2021,Zhang2002,Li2017wide}.

Fluoroethylene carbonate (FEC) is one of the most widely studied electrolyte additives for carbonate-based Li-ion systems. It influences both bulk solvation and interfacial chemistry and can promote LiF-rich interphases and improved electrochemical performance when used at appropriate concentrations \cite{Zhang2021,Starovoytov2021,Yao2022,Brown2018,Hou2021,Xu2019,Lin2020,Hou2019}. At the same time, excessive FEC can increase viscosity and strengthen local coordination, so its effect is not monotonic and depends on both solvent composition and temperature \cite{Yang2023electrolyte,He2022}. For that reason, the key question is not whether FEC is beneficial in general, but under which compositional window it improves transport without overconstraining the Li$^+$ solvation shell. Because these interfacial benefits can coexist with markedly different bulk solvation and transport behavior, FEC provides a useful test case for separating bulk-transport effects from broader, interfacially mediated aspects of battery performance; the present study isolates the former.

Atomistic simulation is well suited to address this question. MD simulations can resolve temperature-dependent diffusion, ion pairing and solvent coordination in ways that are difficult to isolate experimentally \cite{Yoo2023,Chen2024,deKlerk2018}. In particular, radial distribution function (RDF) and coordination-number (CN) analyses can directly connect transport trends to local structural rearrangements around Li$^+$.  Two prior studies are especially relevant: Ringsby~\textit{et al.} characterized low-temperature transport in LP57-class EC/EMC electrolytes \cite{Ringsby2021}, while Hou~\textit{et al.} examined EC/EMC/FEC solvation and transport at ambient conditions \cite{Hou2021}. What remains unresolved is the coupled dependence of low-temperature bulk transport on solvent composition and FEC loading: in particular, whether the apparent benefits of low FEC concentrations persist once collective ion- ion correlations are included in the conductivity, and whether the transport response is governed primarily by solvent-matrix mobility or by changes in Li$^+$ solvation and ion pairing. The present work addresses this gap by sweeping the FEC composition axis systematically, evaluating both the NE and GK conductivity definitions at every state point, and coupling the resulting dataset to a machine-learning surrogate layer.

In this work, we investigate 1~M LiPF$_6$ electrolytes in pure EC, pure EMC, and EC/EMC (3:7, w/w), together with EC and EC/EMC formulations containing 1, 2, 5 and 10~mol\% FEC. Simulations were carried out from 298 to 233~K to quantify the coupled effects of temperature, solvent composition and additive loading on Li$^+$ diffusion, ionic conductivity and local solvation structure. The main objective is to identify the FEC concentration regime that best preserves low-temperature transport and to clarify the structural mechanisms responsible for that behavior.

Although MD provides mechanistic resolution, exhaustive exploration of composition-temperature space is computationally expensive because each new electrolyte requires separate equilibration and production trajectories.  The conductivity trends obtained here vary systematically with solvent composition, FEC loading and temperature, motivating an interpolative surrogate model within the sampled domain. We therefore augment the atomistic analysis with a machine-learning regression layer trained directly on the MD outputs. Machine-learning-assisted electrolyte screening is an active field spanning simulation- and experiment-driven approaches \cite{shi2025,chang2025}; here, composition-disjoint nested cross-validation is used so that reported errors refer to held-out formulations, while feature attribution is treated as a description of the fitted surrogate rather than as independent physical evidence. The purpose of this layer is not to replace MD outside the simulated domain, but to accelerate screening within the studied formulation window by rapidly estimating conductivity at unsimulated composition-temperature conditions.

\section{Methods}

\subsection{Molecular Dynamics Simulations}

Classical MD simulations were performed with LAMMPS (Large-scale Atomic/Molecular Massively Parallel Simulator) \cite{Plimpton1995}. Force-field parameters for EC, EMC and FEC were generated with LigParGen using OPLS-AA-compatible parametrization \cite{Dodda2017,Jorgensen2005,Dodda2017charge}. Parameters for PF$_6^-$ and Li$^+$ were taken from the literature \cite{CanongiaLopes2004,Jensen2006}. A nonpolarizable force field was used throughout.  Bonded interactions (bond stretching, angle bending and improper terms) were described with harmonic functional forms and dihedral rotations with the OPLS torsional potential, while nonbonded interactions were treated with 12-6 Lennard-Jones and Coulomb potentials using geometric mixing rules, analytic tail corrections and the OPLS 1-4 scaling convention.To account for electronic polarization effects absent from the nonpolarizable model, the ionic charges of Li$^+$ and PF$_6^-$ were uniformly scaled by 0.8 according to the electronic continuum correction (ECC), following established practice for carbonate electrolytes \cite{Leontyev2011,Ringsby2021}.
 In the evaluation of the conductivity, by contrast, the formal integer charges ($z=\pm1$) were used. Long-range electrostatics were evaluated with the particle-particle particle-mesh (PPPM) method using a relative accuracy of $10^{-5}$.

\begin{figure}[H]
    \centering
    \includegraphics[width=0.6\textwidth]{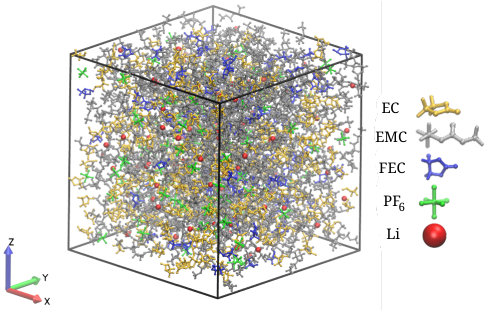}
    \caption{Representative initial simulation configuration generated with PACKMOL for a 1~M LiPF$_6$ EC:EMC electrolyte containing 10~mol\% FEC, visualized using VMD.}
    \label{fig:model}
\end{figure}

Initial configurations were generated with PACKMOL \cite{Martnez2009} in cubic periodic simulation cells. The simulated systems comprised pure EC, pure EMC and EC/EMC (3:7, w/w), as well as EC and EC/EMC (3:7, w/w) with 1, 2, 5 and 10~mol\% FEC. In every case, LiPF$_6$ was added to obtain a nominal salt concentration of 1~M. Because the NVT production volume was set independently at each equilibrated temperature, the instantaneous number-per-volume molarity varies slightly with thermal expansion; ``1~M'' therefore denotes the nominal formulation rather than an exactly temperature-invariant molarity. The resulting cells contain 72--103 ion pairs (72--75 for the EC family, 95 for the EC:EMC family and 103 for neat EMC) and 10{,}600--15{,}824 atoms in cubic boxes of approximately 55~\AA\ edge length. As an example of the generated simulation cells, Figure~\ref{fig:model} shows a representative 1~M LiPF$_6$ EC:EMC electrolyte containing 10~mol\% FEC, visualized using Visual Molecular Dynamics (VMD) was used to inspect representative structures and trajectories \cite{Humphrey1996}. Finite-size (Yeh--Hummer-type) corrections to the self-diffusion coefficients were not applied. Because the leading correction scales approximately as $\Delta D \propto T/(\eta L)$, it is positive for the present systems and therefore acts primarily as an upward systematic correction to the calculated self-diffusion coefficients. Its magnitude is not expected to be uniform across the electrolyte families because of their different viscosities and temperatures. We therefore treat finite-size effects as a potential contribution to the systematic uncertainty of the absolute diffusion coefficients, while noting that the correction is not expected to qualitatively alter the pronounced composition- and temperature-dependent trends reported here.

Each initial configuration was first energy-minimized with the conjugate-gradient algorithm. The minimized systems were then equilibrated for 2~ns in the isothermal isobaric (NPT) ensemble at 298~K and 1~atm with a time step of 1~fs, using the Nos\'e-Hoover thermostat and barostat implemented in LAMMPS (damping constants of 100~fs and 1000~fs, respectively). After this initial equilibration, the systems were cooled from 298 to 233~K in 5~K steps. At each temperature, a 5~ns NPT equilibration run was followed by a 10~ns production run in the canonical (NVT) ensemble at the equilibrated density.  During production, per-species mean-squared displacements were accumulated directly in LAMMPS with center-of-mass drift removal, while the per-species linear momenta required for the ionic charge current were sampled every 2~fs. The direct LAMMPS MSD estimator is trajectory-reference based rather than an explicit many-time-origin average; consequently, diffusion estimates at the slowest state points are treated as sampling-limited and are not used to support fine compositional rankings. VMD was used to inspect representative structures and trajectories. The cooling protocol also does not establish equilibrium liquidus boundaries or nucleation kinetics: a periodic MD cell can remain in a metastable or supercooled condensed state over nanosecond timescales. Accordingly, the lowest-temperature results are interpreted as transport within the simulated liquid-like trajectories and not as evidence that every formulation is an experimentally stable homogeneous liquid at 233~K \cite{Ringsby2021,Su2023}.

\subsection{Transport and Structural Analysis}

 Diffusion coefficients were extracted from the LAMMPS trajectories, and the corresponding ionic conductivity was subsequently calculated using NE relation through Python scripts developed specifically for this analysis. The self-diffusion coefficient of each ionic species was obtained from the long-time mean-squared displacement (MSD) using the Einstein relation
\begin{equation}
D = \frac{1}{6} \lim_{t \rightarrow \infty} \frac{d}{dt}
\left\langle \left|\mathbf{r}(t)-\mathbf{r}(0)\right|^2 \right\rangle .
\end{equation}
To minimize biases associated with ballistic and subdiffusive regimes, the diffusive fitting interval was selected by scanning candidate time windows and identifying the interval whose log- log MSD slope was closest to unity. State points with insufficient diffusive sampling, predominantly at the lowest temperatures of the EMC-containing systems, were flagged in the Supporting Information. Excluding these marginal points changes the fitted Arrhenius activation energies by 5-17\% but does not alter the qualitative separation between the EC- and EMC-containing series. The temperature dependence of the diffusion coefficients was subsequently analyzed using the Arrhenius relation
\begin{equation}
D = D_0 \exp\left(-\frac{E_a}{k_{\mathrm{B}}T}\right),
\end{equation}
where $D_0$ is the pre-exponential factor, $E_a$ is the activation energy, $k_{\mathrm{B}}$ is the Boltzmann constant, and $T$ is the absolute temperature. The activation energy was obtained from the slope of $\ln D$ versus $1/T$.

Ionic conductivity was evaluated using both the NE and GK formalisms. In the NE expression,
\begin{equation}
\sigma_{\mathrm{NE}} =
\frac{e^2}{V k_{\mathrm{B}} T}
\sum_i z_i^2 N_i D_i ,
\end{equation}
where $e$ is the elementary charge, $V$ is the system volume, $z_i$ is the ionic charge number, $N_i$ is the number of ions of species $i$, and $D_i$ is the corresponding self-diffusion coefficient. The formal ionic charges $z_{\mathrm{Li}}=+1$ and $z_{\mathrm{PF_6}}=-1$ were used in the transport operator, whereas the ECC-scaled charges were retained in the interaction potential. This separation follows the two-surfaces protocol of Bl\'azquez \textit{et al.} \cite{Blazquez2023}, in which the scaled charge represents the screened ion-solvent interaction on the potential-energy surface, while the formal charge enters the current operator. The integer-charge prescription is also consistent with the description of charge transport in electronically insulating liquids \cite{GrasselliBaroni2019}; its application here to the present classical fixed-charge force field is regarded as a physically motivated prescription rather than a first-principles derivation. Because both the NE and GK conductivities scale with the square of a common charge rescaling, this convention affects their absolute magnitude but does not alter the corresponding Haven ratio or relative transport trends. The adopted scaling factor of $\zeta=0.8$ has also been independently benchmarked against reference free-ion populations for LiPF$_6$ in EC:EMC \cite{Lehnert2025}. The apparent Li$^+$ transference number was estimated within the same independent-ion approximation as
\begin{equation}
t_{\mathrm{Li}} =
\frac{D_{\mathrm{Li}}}
{D_{\mathrm{Li}}+D_{\mathrm{PF_6}}},
\end{equation}
and is therefore interpreted as a self-diffusion-based mobility estimate rather than as a rigorous electrochemical transference number.

The GK conductivity accounts explicitly for interionic velocity correlations and was calculated from the total ionic charge-current autocorrelation function,
\begin{equation}
\sigma_{\mathrm{GK}} =
\frac{1}{3 V k_{\mathrm{B}} T}
\int_0^\infty
\left\langle \mathbf{J}(0)\cdot\mathbf{J}(t) \right\rangle
dt ,
\end{equation}
where
\begin{equation}
\mathbf{J}(t)=\sum_i z_i e\,\mathbf{v}_i(t)
\end{equation}
is the total ionic charge current. The integral was evaluated using the STable AutoCorrelation Integral Estimator (STACIE) \cite{Toraman2025}, which determines the zero-frequency contribution from the power spectrum and automatically selects the low-frequency fitting range using an information criterion, thereby avoiding an arbitrary correlation-time cutoff. Conductivity uncertainties and quality-control indicators were evaluated for each state point and are reported in the Supporting Information.

Because the complete composition-temperature grid comprises 154 state points, a single production trajectory was generated for each state point rather than multiple independent replicas. Accordingly, the primary emphasis is placed on comparative temperature and composition trends rather than on absolute conductivity values. Within-trajectory statistical uncertainties from the Einstein fits and STACIE estimator were propagated to the reported transport quantities, while state points with limited statistical quality are identified in Table~S1 of the Supporting Information. These uncertainties do not capture between-trajectory variability from independent replicas. In particular, small differences among FEC loadings should not be interpreted as statistically resolved unless they substantially exceed the reported within-trajectory uncertainties; the slowest low-temperature diffusion points are used primarily to establish order-of-magnitude trends.

The degree of ion-motion correlation was quantified using the Haven ratio,
\begin{equation}
H_R=\frac{\sigma_{\mathrm{NE}}}{\sigma_{\mathrm{GK}}},
\end{equation}
with its inverse,
\begin{equation}
\alpha=\frac{1}{H_R}
=\frac{\sigma_{\mathrm{GK}}}{\sigma_{\mathrm{NE}}},
\end{equation}
representing the ionicity. Values of $\alpha<1$ indicate suppression of charge transport relative to the independent-ion limit due to correlated ionic motion, whereas $\alpha\approx1$ corresponds to nearly uncorrelated transport.

Local solvation structure was characterized from radial distribution functions (RDFs) and coordination numbers (CNs). RDFs were calculated from the MD trajectories using VMD. The RDF between atoms $x$ and $y$ is defined as
\begin{equation}
g(r) = \frac{n(r)}{4\pi r^2 \rho\,dr},
\end{equation}
where $n(r)$ is the number of $y$ atoms within a spherical shell of thickness $dr$ at distance $r$ from atom $x$, and $\rho$ is the bulk number density of species $y$. The corresponding coordination number was obtained by integrating the RDF up to its first minimum,
\begin{equation}
N(r) = 4\pi \rho \int_0^r r^2 g(r)\,dr .
\end{equation}
Particular attention was given to Li$^+$-O correlations involving the carbonyl oxygens of EC and EMC and to Li$^+$-F and Li$^+$-P correlations involving PF$_6^-$, as these provide complementary measures of first-shell solvation and ion association.

\subsection{Machine-Learning Surrogate Modeling}

To convert the MD dataset into a fast electrolyte-screening tool, surrogate models were trained to predict the total NE conductivity, $\sigma_{\mathrm{NE}}$, and the total GK conductivity, $\sigma_{\mathrm{GK}}$, from electrolyte composition and temperature. The supervised-learning dataset consisted of all 154 MD state points, spanning 11 composition groups over 14 temperature points; no state points were excluded. Each sample was represented by the renormalized solvent mole fractions $x_{\mathrm{EC}}$, $x_{\mathrm{EMC}}$ and $x_{\mathrm{FEC}}$ (with $x_{\mathrm{EC}}+x_{\mathrm{EMC}}+x_{\mathrm{FEC}}=1$), the temperature $T$, the inverse temperature $1000/T$, a binary indicator for FEC presence and composition temperature interaction terms of the form $x_i\,(1000/T)$. Two feature settings were examined: a numeric-only representation and a representation augmented with a categorical solvent-family label (EC, EC:EMC or EMC). The descriptor set deliberately retains collinear terms ($T$ and $1000/T$, and the compositionally constrained mole fractions), which the nonlinear learners tolerate; the feature-attribution analysis later shows that the categorical solvent-family label adds only minor information beyond the continuous composition fractions. One consequence of retaining deterministically related descriptors is that the permutation-based SHAP explainer necessarily queries the fitted model at feature combinations that violate the built-in constraints (mole fractions summing to unity, $1000/T$ consistent with $T$, interaction terms consistent with their factors); the reported attributions are therefore read as a description of what the fitted function relies on rather than as a strictly causal decomposition.

Five regression families were benchmarked: polynomial ridge regression, support-vector regression (SVR) with a radial-basis kernel, Gaussian process regression (GPR), random forest and extreme gradient-boosted trees (XGBoost) \cite{chen2016}. Numeric features were median-imputed and, for the ridge, support-vector and Gaussian-process models, standardized; the ridge model additionally used a degree-2 polynomial feature expansion, and the categorical solvent-family label was one-hot encoded. For each conductivity target, both the untransformed response and a $\log(1+y)$ target transform were evaluated to accommodate the strongly right-skewed, near-zero values reached at low temperature. Separate single-target models were trained for $\sigma_{\mathrm{NE}}$ and $\sigma_{\mathrm{GK}}$, and multi-output variants were also tested to determine whether joint learning of the two conductivity definitions improves predictive performance. Hyperparameters were selected by nested group-aware cross-validation, using composition group as the grouping variable so that all temperatures belonging to a given composition were assigned to the same fold; within each inner partition, model hyperparameters were tuned by exhaustive grid search scored on RMSE. This design makes the test folds composition-disjoint and therefore probes generalization to unseen electrolyte formulations rather than only to unseen temperatures. The outer loop used five folds and the inner loop used up to four folds depending on the number of composition groups available in the training partition. A full leave-one-composition-out design (eleven outer folds, one per composition group) would probe per-composition generalization at finer granularity, at the cost of appreciably smaller and less balanced training partitions in each fold; the five-fold grouping was chosen to keep the training partitions large enough for stable hyperparameter selection, and the generalization claims reported below should be read at that granularity. Model quality was assessed with the root-mean-square error (RMSE), mean absolute error (MAE) and coefficient of determination ($R^2$), with the mean absolute percentage error (MAPE) retained only as a diagnostic of the low-conductivity regime. All models were implemented in scikit-learn \cite{pedregosa2011} and XGBoost \cite{chen2016} under a fixed random seed (42; scikit-learn 1.8, XGBoost 3.2, SHAP 0.52), and feature attributions of the deployed surrogates were computed with Shapley additive explanations (SHAP) \cite{lundberg2017}.

After model selection, the best-performing surrogates were refit on the full 154-sample dataset and used to generate dense conductivity maps over composition-temperature space. Predictive uncertainty was summarized using empirical intervals derived from out-of-fold residuals, grouped (composition-level) bootstrap resampling and, for untransformed-target variants only, the native Gaussian-process predictive standard deviation. Because multiple temperatures from the same composition are statistically clustered, the residual intervals are treated as cross-validated empirical uncertainty bands rather than as exact split-conformal intervals with distribution-free coverage guarantees \cite{lei2018}. The dataset, the refit surrogate models and the analysis notebook are provided to support full reproducibility.

\section{Results and Discussion}

\begin{figure}[H]
    \centering
    \includegraphics[width=0.70\textwidth]{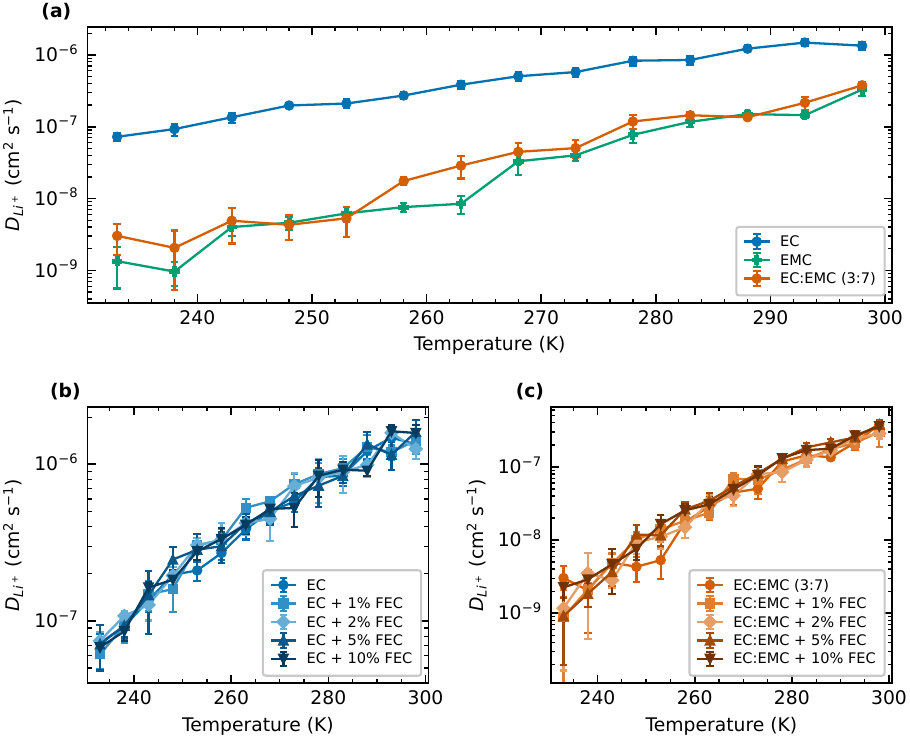}
    \caption{Temperature dependence of the Li$^+$ diffusion coefficient in 1~M LiPF$_6$ electrolytes: (a) the three base solvents EC, EMC and EC/EMC (3:7, w/w); (b) EC with 0, 1, 2, 5 and 10~mol\% FEC; (c) EC/EMC (3:7, w/w) with 0, 1, 2, 5 and 10~mol\% FEC. Error bars denote one standard deviation of the windowed Einstein fit.}
    \label{fig:diff_all}
\end{figure}

We first examine the effect of temperature on Li$^+$ diffusion in the three base electrolytes: neat EC, neat EMC and EC/EMC (3:7, w/w), all containing 1~M LiPF$_6$. As shown in Figure~\ref{fig:diff_all}a, the Li$^+$ diffusion coefficient decreases monotonically with decreasing temperature in all three systems, consistent with progressively slower structural relaxation and solvent-mediated transport at low temperature. The three electrolytes, however, do not respond identically. The EC system retains comparatively high Li$^+$ mobility over the full temperature range: $D_{\mathrm{Li}}$ decreases from $1.34\times10^{-6}$~cm$^2$~s$^{-1}$ at 298~K to $7.2\times10^{-8}$~cm$^2$~s$^{-1}$ at 233~K, a factor of about 19. This stability is consistent with strong Li$^+$ solvation and a relatively stable first-shell coordination environment \cite{Ong2015}. The EMC system shows a far more pronounced loss of Li$^+$ diffusion upon cooling, from $3.3\times10^{-7}$ to $1.3\times10^{-9}$~cm$^2$~s$^{-1}$ (a factor of about 250), despite its lower viscosity at ambient temperature. This indicates that low bulk viscosity alone does not guarantee better cold transport; weaker Li$^+$ solvation and a larger PF$_6^-$ first-shell population can offset the nominal viscosity advantage \cite{Liu2020}. We note that viscosity itself was not computed in this work; the viscosity comparisons invoked here and in the Introduction reflect established experimental behavior of these solvents rather than values derived from the present trajectories. The EC/EMC mixture tracks the EMC-like response ($D_{\mathrm{Li}}$ falls by a factor of about 125 over the same interval), showing that the mixed carbonate has much stronger temperature sensitivity than the neat-EC simulation over the sampled range.

The influence of FEC on transport becomes clearer when the EC and EC/EMC systems are considered separately. For EC-based electrolytes (Figure~\ref{fig:diff_all}b), all compositions show the expected decline in diffusion with decreasing temperature, and the five curves remain within a factor of about two of one another over the entire range. The FEC effect is therefore a modulation rather than a qualitative change. In the intermediate cold range (approximately 273-248~K) the 1~mol\% FEC system shows consistently higher Li$^+$ diffusion than neat EC (positive at all five shared temperatures; pooled difference of 2.6 standard errors; for example, $3.0\times10^{-7}$ versus $2.1\times10^{-7}$~cm$^2$~s$^{-1}$ at 253~K), with a weaker tendency in the same direction at 2~mol\%.  FEC can alter Li$^+$ solvation even as a minority species \cite{Hou2019}; in the present trajectories, 1-2~mol\% FEC is associated with a modest mobility increase at intermediate temperatures, but the mechanism cannot be assigned directly to FEC first-shell coordination because Li$^+$-FEC RDFs were not evaluated. At the cold limit of 233~K the differences between compositions contract to within the fit uncertainties. The 5 and 10~mol\% systems remain competitive in single-ion diffusion; as shown below, their penalty appears more clearly in the collective (GK) conductivity than in $D_{\mathrm{Li}}$ itself.

The EC/EMC family (Figure~\ref{fig:diff_all}c) behaves differently. Because the mixed-carbonate matrix itself loses two orders of magnitude in $D_{\mathrm{Li}}$ on cooling, the FEC modulation is small compared with the solvent effect, and below about 253~K the compositional differences are within the statistical uncertainties of the individual state points. In this family the additive neither rescues nor systematically degrades the low-temperature diffusivity; the collapse is governed by the EMC-rich solvent matrix. Taken together, panels b and c of Figure~\ref{fig:diff_all} indicate that the first-order lever on the simulated cold-temperature diffusion response is the base solvent (EC-rich versus EMC-rich), while 1-2~mol\% FEC shows a smaller, suggestive enhancement in the EC family at intermediate temperatures that requires independent-replica confirmation.

\begin{figure}[H]
    \centering
    \includegraphics[width=0.70\textwidth]{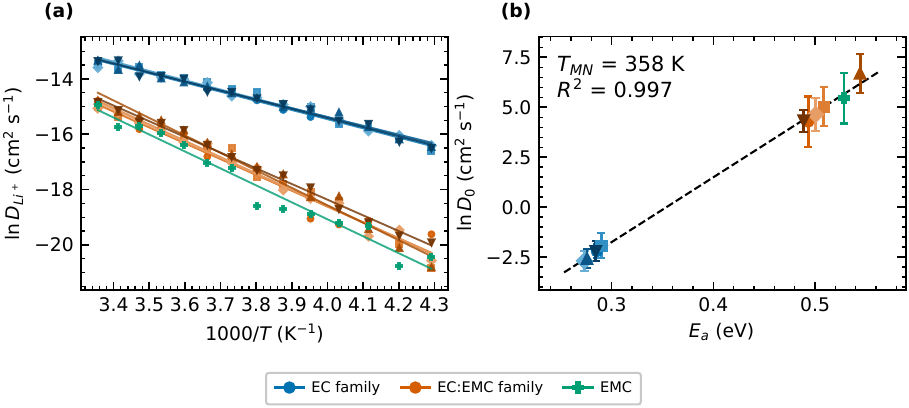}
    \caption{(a) Arrhenius plots of the Li$^+$ diffusion coefficient for all eleven electrolytes with linear fits. (b) Meyer-Neldel compensation plot of the pre-exponential factor $D_0$ versus activation energy $E_a$; the line is a linear fit to $\ln D_0$ versus $E_a$ ($R^2=0.997$), corresponding to a compensation temperature $T_{\mathrm{MN}}\approx358$~K.}
    \label{fig:arrhenius}
\end{figure}

The Arrhenius parameters obtained from the Li$^+$ diffusion data are summarized in Figure~\ref{fig:arrhenius}, with the corresponding fits shown alongside. The Meyer-Neldel analysis is included as a descriptive compensation diagnostic rather than as a central mechanistic result. Alternative temperature-dependent descriptions were also considered to assess the robustness of the Arrhenius representation; however, over the 65~K temperature window sampled here, the Arrhenius model provides an adequate and statistically competitive description of the diffusion data. We therefore retain the Arrhenius description throughout. The eleven electrolytes separate cleanly into two kinetic groups. The EC-based systems (neat EC and EC with 1-10~mol\% FEC) exhibit activation energies of $E_a=0.273$-0.290~eV, whereas all EMC-containing systems (neat EMC, EC/EMC and EC/EMC with FEC) fall between 0.489 and 0.544~eV (uncertainty-weighted fit standard errors of $\pm0.01$-$0.03$~eV), almost a factor of two higher. This separation is robust to the low-temperature sampling limitation noted in Methods: excluding the 35 marginally subdiffusive state points (all in EMC-containing systems at the lowest temperatures) shifts the fitted activation energies by 5-17\%, but the twofold gap between the EC and EMC-containing families persists under this filtering. FEC loading shifts $E_a$ only weakly within each family (by at most 0.06~eV), showing that the fitted barrier is much more sensitive to host-solvent family than to FEC loading; this observation alone does not establish whether the microscopic transport pathway is reorganized. The lowest barrier in the data set belongs to EC with 2~mol\% FEC ($E_a=0.273$~eV), closely followed by the other EC-family compositions. For reference, Rampal \textit{et al.} \cite{rampal2024structural} reported $E_a=0.55$~eV for a comparable 1~M LiPF$_6$/EC system with unscaled ionic charges; the lower barriers found here are consistent with the faster, less ion-paired dynamics expected from the electronic-continuum charge scaling employed in this work.

\begin{figure}[H]
\centering
\includegraphics[width=0.70\textwidth]{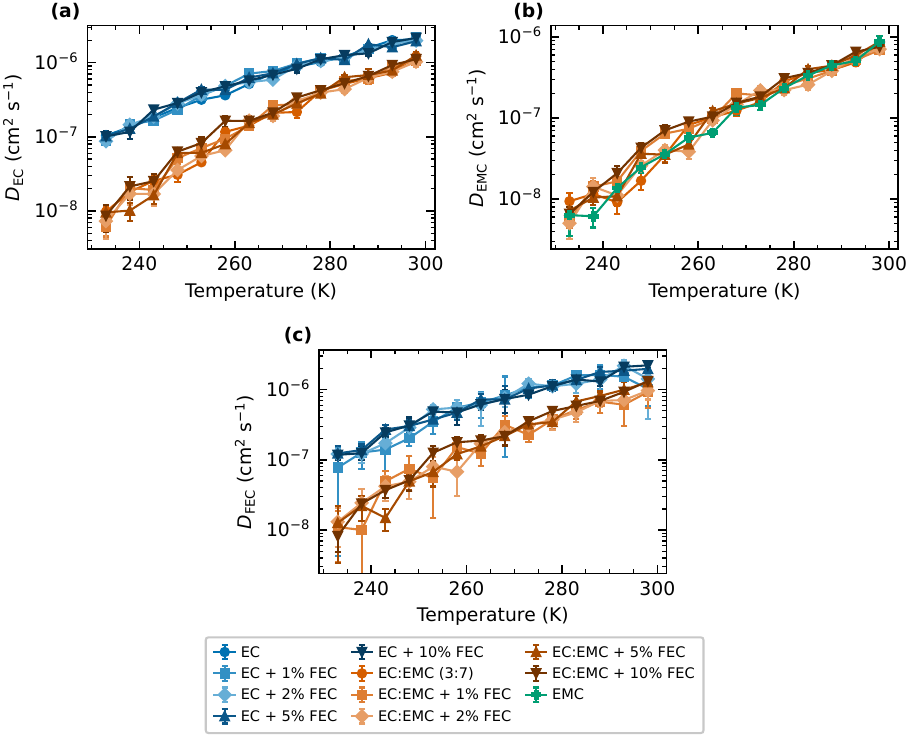}
\caption{Temperature dependence of the solvent self-diffusion coefficients in the simulated 1~M LiPF$_6$ electrolytes: (a) EC, (b) EMC and (c) FEC. Each species is shown for every composition in which it is present; colors follow the electrolyte color coding used throughout. Error bars denote one standard deviation of the windowed Einstein fit.}
\label{fig:solvent_diff}
\end{figure}

The distinct transport behavior of the EC- and EMC-containing systems can be further examined in terms of solvent dynamics, as Li$^+$ migrates within a continuously evolving solvation environment whose mobility provides an important dynamic background for ion transport. Figure~\ref{fig:solvent_diff} shows the self-diffusion coefficients of EC, EMC, and FEC in all systems where the respective species is present.

\noindent The solvent molecules are consistently more mobile than Li$^+$: at 298~K the EC self-diffusion coefficient exceeds $D_{\mathrm{Li}}$ by a factor of about 1.4-1.6 in the EC family and by a factor of 3.1-3.6 in the EC/EMC family, and the gap widens on cooling in the mixed carbonate (factors of 5-8.5 at 253~K). A solvent-to-cation diffusivity ratio above unity is consistent with a substantial vehicular contribution, although this ratio alone does not establish a persistent-solvation-shell mechanism; the same hierarchy $D_{\mathrm{solvent}}>D_{\mathrm{PF_6}}>D_{\mathrm{Li}}$ is observed experimentally in NMR measurements on LiPF$*6$ carbonate electrolytes \cite{Hayamizu2012} and in earlier simulation work on carbonate blends \cite{Borodin2009}. PFG-NMR measurements on 1~M LiPF$*6$ in EC:DEC report the identical ordering $D*{\mathrm{EC}}\approx D*{\mathrm{DEC}}>D_{\mathrm{PF_6}}>D_{\mathrm{Li}}$ \cite{Feng2017}, and a combined MD-eNMR study of binary carbonate blends likewise concludes that Li$^+$ transport is more vehicular than that of PF$_6^-$ \cite{Lehnert2025}. The larger ratio in the EC/EMC family indicates that on cooling the Li$^+$ ion is slowed disproportionately relative to its solvent matrix, consistent with the increasing importance of local coordination and solvent-coupled transport limitations rather than with a purely viscosity-driven slowdown.

The solvent activation energies provide a complementary view of this coupling. The solvent dynamics likewise separate into the two kinetic groups identified for Li$^+$ transport: Arrhenius fits to the solvent data give $E_a=0.26$-0.29~eV for EC and FEC in the EC family and 0.39-0.46~eV for all solvent species in the EMC-containing systems. The family-level separation in cation activation energy tracks the corresponding separation in solvent activation energy, consistent with an important contribution from solvent-matrix relaxation together with the composition-dependent anion coordination and net ion correlations discussed below. The relation between solvent and cation barriers, however, differs between the groups. In the EC family the Li$^+$ and solvent barriers agree within about 0.02~eV, which is consistent with strongly coupled cation and matrix relaxation but does not by itself prove a unique rate-limiting mechanism. In the EMC-containing systems the Li$^+$ barrier exceeds the solvent barriers by 0.05-0.11~eV, so the cation pays a measurable additional energetic cost beyond matrix relaxation. This additional barrier is consistent with the larger PF$_6^-$ first-shell population observed for the EMC-rich systems in the structural analysis, together with the greater sensitivity of their Li$^+$ transport to cooling.

The mixed-carbonate systems further reveal an asymmetry between the two co-solvents. At 298~K EMC diffuses about 33\% slower than EC in the same electrolyte, even though neat EMC is the lower-viscosity solvent; by 233~K the two diffusivities converge. The present RDF coordination counts do not support assigning this asymmetry to preferential EC binding: under the atom-counting convention used here, the reported Li-O$_{\mathrm{EMC}}$ coordination numbers are numerically much larger than the Li-O$_{\mathrm{EC}}$ values. The solvent-diffusion asymmetry is therefore reported as an observation rather than as evidence that EC dominates the first shell. FEC tracks the diffusivity of its host solvents at all compositions and temperatures, indicating that no distinct slow FEC population is resolved by the self-diffusion data. Because direct Li-O$_{\mathrm{FEC}}$ coordination was not included in the present RDF analysis, these data do not by themselves establish how FEC enters or exchanges within the Li$^+$ first shell.

\begin{figure}[H]
    \centering
    \includegraphics[width=0.70\textwidth]{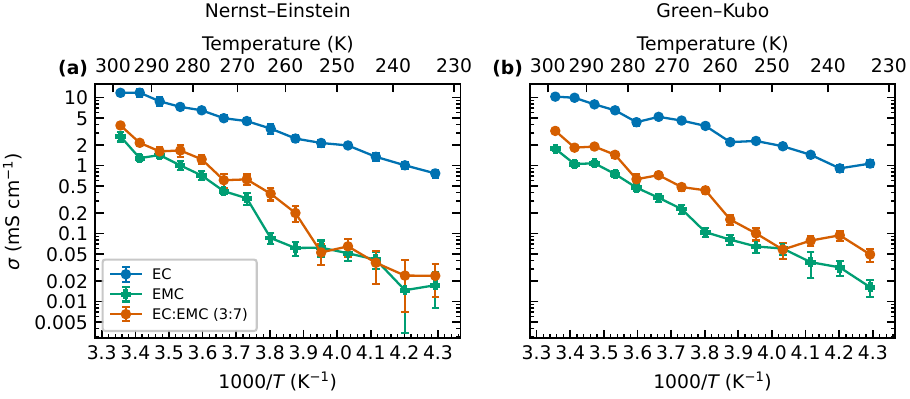}
    \caption{(a) NE and (b) GK ionic conductivities for 1~M LiPF$_6$ in EC, EMC and EC/EMC (3:7, w/w), plotted against inverse temperature. Error bars denote one standard deviation; GK values and uncertainties are obtained with the STACIE spectral estimator.}
    \label{fig:cond_base}
\end{figure}

To relate the diffusion trends to charge transport, we evaluated the conductivity using both NE and GK approaches. The GK values, which account for all interionic correlations, constitute the primary conductivity results of this work, while NE serves as an independent-ion reference. Figure~\ref{fig:cond_base} shows the NE (Figure~\ref{fig:cond_base}a) and GK (Figure~\ref{fig:cond_base}b) conductivities of the three base electrolytes against inverse temperature; both estimators increase with temperature, as expected. The EC system exhibits the highest conductivity at 298~K, with $\sigma_{\mathrm{GK}}=10.29\pm0.49$~mS~cm$^{-1}$ ($\sigma_{\mathrm{NE}}=11.71\pm1.14$~mS~cm$^{-1}$). EMC shows the lowest values, with $\sigma_{\mathrm{GK}}=1.76\pm0.14$~mS~cm$^{-1}$ ($\sigma_{\mathrm{NE}}=2.67\pm0.42$~mS~cm$^{-1}$) at 298~K, while EC/EMC remains intermediate at $\sigma_{\mathrm{GK}}=3.22\pm0.24$~mS~cm$^{-1}$. The neat-EMC value agrees closely with an independent OPLS-AA simulation of the same electrolyte, which reports a NE conductivity of 1.58~mS~cm$^{-1}$ at 300~K \cite{Kasemchainan2025}; both simulated values lie a factor of 2-3 below the measured conductivity of about 4~mS~cm$^{-1}$ for 1~M LiPF$_6$ in EMC at 298~K \cite{Xiong2018}. For reference, the measured room-temperature conductivity of the corresponding commercial LP57 formulation (1~M LiPF$_6$ in EC:EMC 3:7, w/w) is approximately 9~mS~cm$^{-1}$ \cite{Landesfeind2019}, and an independent electrochemical characterization of the same solvent ratio reports 9.5~mS~cm$^{-1}$ at the 1~M conductivity maximum \cite{Nyman2008}, about a factor of three above the present GK value. Importantly, the disagreement is not purely a common multiplicative offset: the present force field ranks neat EC substantially above EC:EMC at 298~K, whereas experimental LP57 conductivity is comparable to typical EC-based carbonate values \cite{Hou2021,Landesfeind2019}. This qualitative ranking discrepancy, together with the absolute underprediction for LP57, limits transferability of the simulated ranking to experiment and likely reflects limitations of the classical force field and charge-treatment protocol. The calculated conductivities are therefore interpreted as trends internal to the present model rather than as quantitatively calibrated predictions. This ranking highlights an important point: bulk conductivity is governed not only by single-ion diffusion, but also by ion dissociation and cross-correlated motion. The same decoupling is observed experimentally: Landesfeind and Gasteiger measured nearly indistinguishable salt diffusion coefficients for EC:DMC, EC:EMC and EMC:FEC electrolytes despite pronounced differences in their ionic conductivities, and attributed the divergence to differing degrees of ion association rather than to ionic mobility \cite{Landesfeind2019}. A solvent with lower viscosity is not automatically the best conductor if, at a given temperature, it also promotes stronger ion pairing. The contrast sharpens dramatically on cooling: at 233~K the EC system still delivers $\sigma_{\mathrm{GK}}=1.07$~mS~cm$^{-1}$, about 10\% of its room-temperature value, whereas EC/EMC and EMC collapse to 0.049 and 0.016~mS~cm$^{-1}$, respectively, retaining less than 2\% of their 298~K conductivity; because these two values approach the resolution floor of the conductivity estimators at this temperature, they are best read as evidence of an order-of-magnitude suppression relative to EC rather than as precise absolute figures.

\begin{figure}[H]
    \centering
    \includegraphics[width=0.70\textwidth]{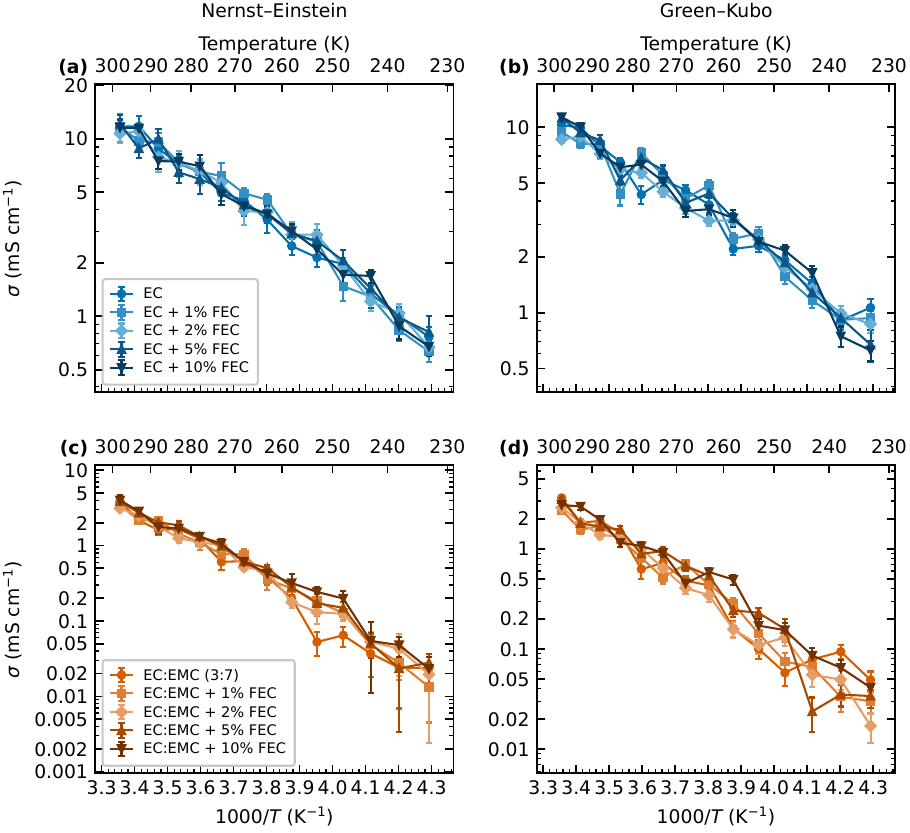}
    \caption{Ionic conductivities of the FEC-containing electrolyte families plotted against $1000/T$: (a) Nernst-Einstein (NE) and (b) Green-Kubo (GK) conductivities for 1~M LiPF$_6$/EC with 0, 1, 2, 5, and 10~mol\% FEC; (c) NE and (d) GK conductivities for 1~M LiPF$_6$/EC:EMC (3:7, w/w) with the same FEC loadings. Conductivity is plotted on a logarithmic scale, with the corresponding temperature shown on the secondary top axes. Error bars denote one standard deviation.
}
    \label{fig:cond_families}
\end{figure}

For EC-based systems (Figure~\ref{fig:cond_families}a,b), the FEC effect at room temperature is modest: all five compositions lie in a band of $\sigma_{\mathrm{GK}}=8.6$-11.3~mS~cm$^{-1}$ at 298~K, with the largest NE conductivity in the data set obtained for 5~mol\% FEC ($\sigma_{\mathrm{NE}}=12.12\pm1.49$~mS~cm$^{-1}$). The compositional differentiation emerges at low temperature and separates the family into two regimes. The low-FEC compositions (0-2~mol\%) essentially share one cold-temperature response: their 233~K GK conductivities (0.87-1.07~mS~cm$^{-1}$) all correspond to retaining about 10\% of the respective room-temperature values, and in the intermediate range 273-253~K the 1~mol\% system trends above neat EC (for example, 2.68 versus 2.30~mS~cm$^{-1}$ at 253~K), although at the conductivity level this enhancement remains within the propagated uncertainties (pooled difference $+0.31\pm0.18$~mS~cm$^{-1}$ across the five shared temperatures) and is resolved more clearly in the single-ion diffusion data above. At 233~K, the 5 and 10~mol\% systems end at 0.67 and 0.63~mS~cm$^{-1}$, corresponding to only 5.5-6\% retention and lower values than the 0-2~mol\% group. Between 253 and 243~K, however, the ordering is not monotonic, and several differences are comparable to the propagated uncertainties; the most defensible high-FEC penalty is therefore the reduced retention at the coldest simulated temperature rather than a resolved onset below 253~K. No EC-only RDF analysis is presented here, so this transport trend is not assigned to the EMC-coordination changes observed later for the separate EC:EMC family.

The EC/EMC systems with FEC (Figure~\ref{fig:cond_families}c,d) reproduce the pattern already seen in diffusion: the solvent matrix, not the additive, controls the response. All five compositions start from $\sigma_{\mathrm{GK}}=2.4$-3.2~mS~cm$^{-1}$ at 298~K and converge to 0.017-0.049~mS~cm$^{-1}$ at 233~K, where the remaining compositional differences are comparable to the statistical uncertainties of the individual state points. Within this family the NE and GK estimates track each other closely at moderate temperatures, indicating moderate net ion-correlation effects; the self-diffusion-based mobility ratios are discussed separately below. The robust conclusion is that in EMC-rich matrices FEC cannot compensate the intrinsic low-temperature fragility of the solvent. The near-neutral effect of FEC on room-temperature conductivity in this family agrees with Hou et al., who found nearly equal conductivities for 1.2~M LiPF$_6$ in EC:EMC (3:7) with and without 10\% FEC, in both simulation and experiment \cite{Hou2021}. The host dependence of the additive is also consistent with experiment: in EC-free EMC electrolytes, where ion pairing contributes strongly to the transport response, FEC addition raises the measured conductivity substantially \cite{Xiong2018}, whereas in hosts that already contain a strongly dissociating cyclic carbonate, any additional ion-pair-suppressing effect of FEC is expected to be less pronounced and its net transport effect is correspondingly smaller. Consistent with this picture, when FEC is used at high loading as a full replacement for EC, it can carry the dissociating role on its own: 1~M LiPF$_6$ in FEC:EMC (3:7) reaches a measured conductivity of 8.4~mS~cm$^{-1}$ at 293~K, within the typical range of the EC-based LP57 formulation \cite{StokesRodriguez2025}.

\begin{figure}[H]
    \centering
    \includegraphics[width=0.70\textwidth]{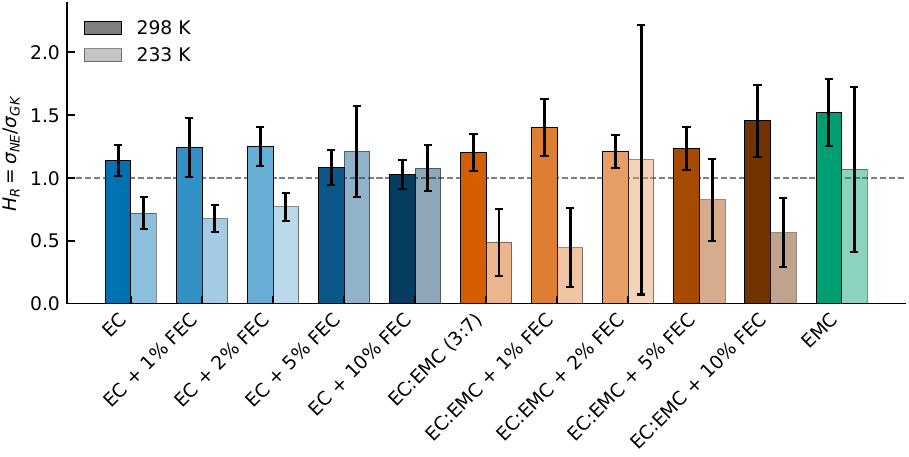}
    \caption{Haven ratios, $H_R=\sigma_{\mathrm{NE}}/\sigma_{\mathrm{GK}}$, at 298 and 233~K for the simulated electrolytes. The dashed line marks the uncorrelated-transport limit $H_R=1$. The 233~K values are shown with reduced emphasis because both estimators become data-limited at the lowest conductivities.}
    \label{fig:haven}
\end{figure}

The systematic offset between the NE and GK estimators observed above is quantified by the Haven ratio $H_R=\sigma_{\mathrm{NE}}/\sigma_{\mathrm{GK}}$ (equivalently the ionicity $\alpha=1/H_R$; note that the inverse convention $\sigma_{\mathrm{GK}}/\sigma_{\mathrm{NE}}$ is also used in the literature), which provides a compact measure of correlated ion motion: $H_R>1$ ($\alpha<1$) indicates that cation-anion correlations suppress charge transport below the independent-ion limit \cite{Nurnberg2022Superionicity}. At 298~K (Figure~\ref{fig:haven}), all eleven electrolytes show $H_R\ge1$, i.e., ionicities between 0.66 and 0.98. No individual composition exceeds the uncorrelated-transport limit by more than two propagated standard deviations, but all eleven compositions fall on the same side of it ($H_R\ge1$). Because these formulations share the same solvent families and FEC loadings, they are not statistically independent; the one-sided pattern is therefore read only as a qualitative indication that net ion-ion correlations tend to suppress the effective conductivity in the present model. The correlation penalty is smallest in the EC family ($\alpha=0.80$-0.98, with the high-FEC compositions closest to unity) and largest in neat EMC ($\alpha=0.66$), which is consistent with the strong Li$^+$-PF$_6^-$ association identified for EMC in the structural analysis. The EC/EMC family occupies an intermediate band ($\alpha=0.69$-0.83). At 233~K the apparent $H_R$ values scatter considerably more widely and several fall below unity (per-state-point values are tabulated in Table~\ref{tab:si_diffusion}); because the underlying conductivities of the EMC-containing systems there are of order $10^{-2}$~mS~cm$^{-1}$, both estimators approach their statistical resolution limits, and $H_R$ at this temperature is not used for quantitative mechanistic inference. Within the present force field, the higher-conductivity formulations at ambient and moderately low temperatures combine higher self-diffusion with a smaller net correlation penalty. These Haven ratios agree with prior observations on closely related systems: for 1.2~M LiPF$_6$ in EC:EMC (3:7) with and without 10\% FEC, Hou et al. found the NE estimate to overstate the correlated conductivity by more than 40\% ($H_R>1.4$) \cite{Hou2021}, and an NE-to-GK ratio of about 1.4 has likewise been reported for concentrated aqueous alkali-halide solutions \cite{Blazquez2023}, suggesting that a correlation penalty of this magnitude is a generic feature of concentrated 1:1 electrolytes rather than an artifact of a particular force field or solvent. The quantity $\alpha=1/H_R$ is therefore used here only as a compact measure of the net correlation penalty, not as the fraction of dissociated ions: distinct cation-cation, anion-anion and cation-anion correlations contribute to the GK current, and Haven-ratio-based dissociation estimates need not equal concentrated-solution-theory measures \cite{Feng2017}.

The systematic offset between the NE and GK estimators observed above can be quantified by the Haven ratio, $H_R=\sigma_{\mathrm{NE}}/\sigma_{\mathrm{GK}}$, with the reciprocal quantity $\alpha=1/H_R$ sometimes termed the ionicity. Note that the inverse Haven-ratio convention, $\sigma_{\mathrm{GK}}/\sigma_{\mathrm{NE}}$, is also used in the literature. Here, $H_R>1$ ($\alpha<1$) indicates that net ion-ion correlations reduce the ionic conductivity relative to the independent-ion limit \cite{Nurnberg2022Superionicity}. At 298~K (Figure~\ref{fig:haven}), all eleven electrolytes show $H_R\ge1$, corresponding to $\alpha$ values between 0.66 and 0.98. No individual composition differs from the uncorrelated-transport limit by more than two propagated standard deviations, but all eleven compositions fall on the same side of it ($H_R\ge1$). Because these formulations share the same solvent families and FEC loadings, they are not statistically independent; the one-sided pattern is therefore interpreted only as a qualitative indication that net ion-ion correlations tend to suppress the effective conductivity in the present model. The correlation penalty is smallest in the EC family ($\alpha=0.80$-$0.98$), with the high-FEC compositions closest to unity, and largest in neat EMC ($\alpha=0.66$), consistent with the stronger Li$^+$-PF$_6^-$ association identified for EMC in the structural analysis. The EC/EMC family occupies an intermediate range ($\alpha=0.69$-$0.83$). At 233~K, the apparent $H_R$ values scatter considerably more widely, with several falling below unity (per-state-point values are tabulated in Table~\ref{tab:si_diffusion}). Because the underlying conductivities of the EMC-containing systems are of order $10^{-2}$~mS~cm$^{-1}$ at this temperature, both estimators approach their statistical resolution limits; consequently, the 233~K $H_R$ values are not used for quantitative mechanistic inference. Within the present force field, the higher-conductivity formulations at ambient and moderately low temperatures combine higher self-diffusion with a smaller net correlation penalty. The present Haven-ratio trends are consistent with prior observations in closely related systems. For 1.2~M LiPF$_6$ in EC:EMC (3:7) with and without 10\% FEC, Hou et al. found that the NE estimate overstated the correlated conductivity by more than 40\% ($H_R>1.4$) \cite{Hou2021}. An NE-to-GK ratio of about 1.4 has likewise been reported for concentrated aqueous alkali-halide solutions \cite{Blazquez2023}, suggesting that correlation penalties of this magnitude are not unique to a particular force field or solvent. Accordingly, $\alpha=1/H_R$ is used here only as a compact measure of the net correlation penalty, not as a measure of the fraction of dissociated ions. Distinct cation-cation, anion-anion, and cation-anion correlations all contribute to the GK current, and Haven-ratio-based dissociation estimates need not coincide with concentrated-solution-theory measures \cite{Feng2017}.

\begin{figure}[H]
    \centering
    \includegraphics[width=0.70\textwidth]{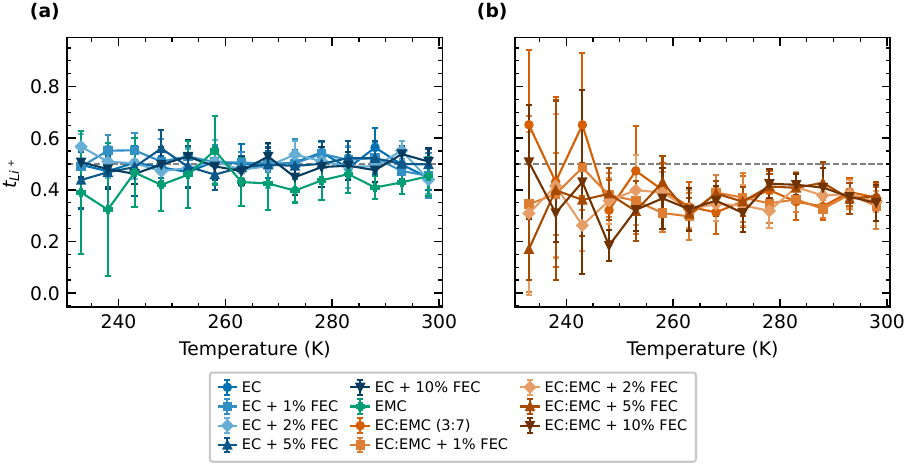}
    \caption{Apparent Li$^+$ transference number $t_{\mathrm{Li}}=D_{\mathrm{Li}}/(D_{\mathrm{Li}}+D_{\mathrm{PF_6}})$ as a function of temperature for the EC-based (a) and EC:EMC-based (b) electrolyte families. The dashed line marks $t_{\mathrm{Li}}=0.5$.}
    \label{fig:transference}
\end{figure}

The Haven ratio quantifies the average correlation penalty on the total ionic current, while a complementary self-diffusion perspective is provided by the apparent Li$^+$ mobility fraction defined in Methods. Figure~\ref{fig:transference} shows values of 0.17-0.65, with the EC-based electrolytes clustering at $t_{\mathrm{Li}}=0.44$-0.51 at 298~K and the EC/EMC family systematically lower at 0.34-0.38. The bounded range $0\le t_{\mathrm{Li}}\le1$ follows by construction from the positive self-diffusion coefficients and therefore does not constitute an independent physical validation. On cooling, these apparent transference values show no strong systematic drift, although the scatter increases at the lowest temperatures, where the diffusion fits carry their largest relative uncertainties. Thus, the low-temperature conductivity loss documented above is more consistent with a reduction in overall ionic mobility than with a selective suppression of cation transport. The value of 0.37 for EC/EMC agrees closely with a combined MD-eNMR study of the same electrolyte (0.373 MD, 0.343 eNMR) \cite{Lehnert2025}; electrochemical polarization measurements yield substantially lower, temperature-dependent values \cite{Landesfeind2019}, emphasizing that the present self-diffusion ratio is not a rigorous electrochemical transference number and is not assumed to provide a general upper bound.

\begin{figure}[H]
    \centering
    \includegraphics[width=0.70\textwidth]{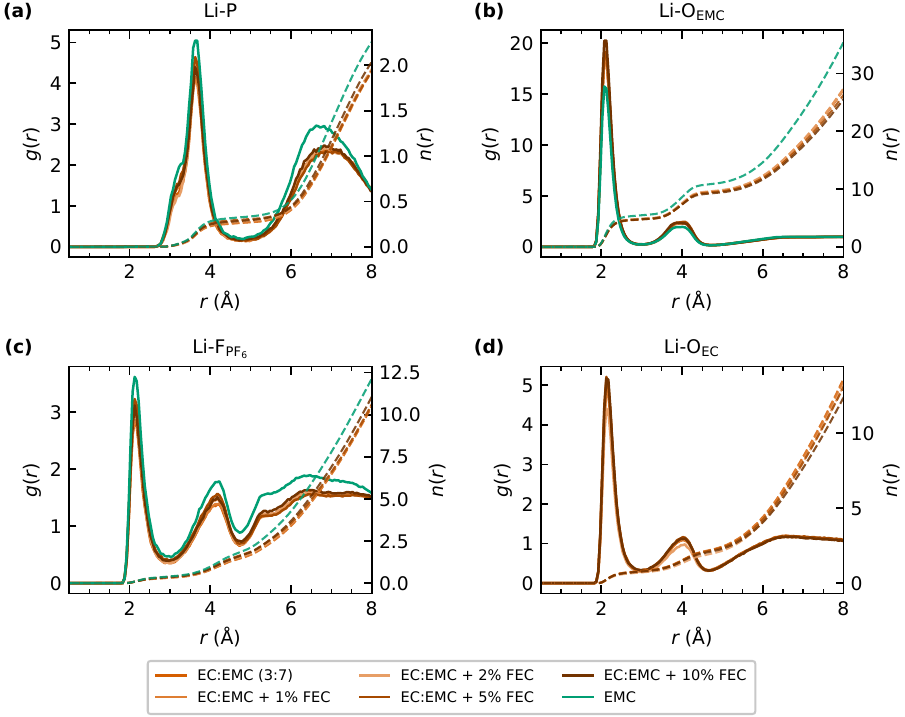}
    \caption{RDF and CN profiles at 298~K for the six selected electrolyte models: (a) Li-P, (b) Li-O$_{\mathrm{EMC}}$, (c) Li-F$_{\mathrm{PF_6}}$ and (d) Li-O$_{\mathrm{EC}}$.}
    \label{fig:rdf_298}
\end{figure}

The transport results above establish clear composition-dependent differences in Li$^+$ mobility, motivating a closer examination of the local coordination environment that may underlie these trends. To characterize the local structural environment of Li$^+$, RDF and CN analyses were performed for Li$^+$-P, Li$^+$-F${\mathrm{PF_6}}$, Li$^+$-O${\mathrm{EMC}}$, and Li$^+$-O${\mathrm{EC}}$ pairs in the EC(3:7)-based family containing 0, 1, 2, 5, and 10~mol\% FEC, together with neat EMC as a solvent reference. Li$^+$-O$_{\mathrm{FEC}}$ correlations were not included in the present structural dataset; therefore, direct FEC occupancy of the Li$^+$ first shell is not quantified and FEC-induced structural effects below are restricted to changes observed in the EC, EMC and PF$_6^-$ coordination channels. At 298~K (Figure~\ref{fig:rdf_298}), the Li-P and Li-F$_{\mathrm{PF_6}}$ correlations indicate that neat EMC exhibits the strongest Li$^+$-PF$_6^-$ association among the six systems. The first Li-P maximum occurs at approximately 3.62-3.68~\AA\ across the systems, while the corresponding coordination number is highest in neat EMC ($N=0.317$). Within the EC:EMC(3:7)-based FEC series, the Li-P coordination number increases from 0.252 at 0~mol\% FEC to 0.297 at 5~mol\% FEC and then decreases slightly to 0.282 at 10~mol\% FEC. A similar trend is observed for the Li-F$_{\mathrm{PF_6}}$ correlation, for which the coordination number increases from 0.299 to 0.342 between 0 and 5~mol\% FEC and decreases slightly to 0.328 at 10~mol\% FEC, while neat EMC exhibits the highest value ($N=0.384$). Thus, both Li-P and Li-F$_{\mathrm{PF_6}}$ correlations indicate a moderate increase in Li$^+$-PF$_6^-$ association with FEC addition up to 5~mol\%, followed by a slight decrease at 10~mol\%. The relatively small Li-P and Li-F$_{\mathrm{PF_6}}$ coordination numbers compared with the solvent coordination numbers further indicate that PF$_6^-$ constitutes only a fraction of the Li$^+$ coordination environment. This solvent-dominated coordination picture is consistent with previous molecular-level studies of carbonate-based LiPF$_6$ electrolytes. Borodin and Smith reported that both EC and DMC participate in the Li$^+$ first solvation shell in mixed EC:DMC electrolytes, with a total first-shell population of approximately 4.3-4.6 solvent and anion species, while increasing the DMC fraction enhances anion coordination and ion association \cite{Borodin2009}. Similarly, first-principles simulations by Ong~\textit{et al.} showed that Li$^+$ preferentially forms a solvent-dominated first coordination shell in EC, EMC, and EC/EMC electrolytes, with PF$_6^-$ entering the first shell only in a subset of configurations \cite{Ong2015}. These studies support the present observation that Li$^+$ remains predominantly coordinated by carbonate oxygen atoms, whereas PF$_6^-$ contributes a smaller but finite fraction of the first-shell coordination.

The Li-O$_{\mathrm{EMC}}$ RDFs further show that EMC remains an important component of the Li$^+$ solvation environment across the investigated compositions. The first Li-O$_{\mathrm{EMC}}$ maximum occurs at approximately 2.08-2.12~\AA\ across the systems. Its intensity does not decrease systematically with FEC addition, remaining nearly unchanged between 0 and 1~mol\% FEC and increasing at higher FEC contents, reaching $g(r)=20.23$ at 10~mol\% FEC. The corresponding coordination number remains within a relatively narrow range of 4.719-4.855 EMC oxygen atoms per Li$^+$ for the EC:EMC(3:7)-based FEC series, indicating that the number of EMC oxygen atoms within the first coordination region is comparatively insensitive to FEC content. The variation in RDF peak height should therefore not be interpreted directly as a proportional change in EMC coordination, because the RDF is normalized with respect to the corresponding bulk oxygen density. This effect is also evident in neat EMC, which exhibits the largest Li-O$_{\mathrm{EMC}}$ coordination number ($N=5.421$) despite a lower RDF peak intensity ($g(r)=15.73$). The persistence of solvent coordination is consistent with the results of Hou~\textit{et al.} for the Gen2 EC:EMC(3:7) electrolyte, for which the total Li$^+$ solvent coordination number was reported to be approximately 4.7, with contributions of approximately 1.8 from EC and 2.9 from EMC \cite{Hou2021}.  The present solvent coordination sum is quantitatively higher than the approximately 4.7 reported by Hou~\textit{et al.}; differences in atom selections, integration cutoffs and force-field structure may contribute, so the comparison is used only to support the qualitative conclusion that solvent oxygens dominate over PF$_6^-$ in the analyzed coordination channels. The Li-O$_{\mathrm{EC}}$ RDFs likewise remain comparatively stable across the EC-containing compositions. The first Li-O$_{\mathrm{EC}}$ maximum occurs at approximately 2.12~\AA\ in all EC-containing systems, while the corresponding coordination number varies only moderately from 0.706 to 0.786. These results indicate that EC remains a detectable component of the Li$^+$ first solvation environment despite variations in EMC and FEC content; they do not support describing EC as the dominant first-shell solvent under the present coordination-number definition. The persistence of solvent coordination and the finite Li$^+$-PF$_6^-$ association observed here are therefore consistent with the broader molecular-level picture established for mixed carbonate LiPF$_6$ electrolytes \cite{Hou2019,Borodin2009,Ong2015}. Taken together, the analyzed EC-, EMC- and PF$_6^-$-based coordination channels change only moderately with FEC loading at 298~K. Because Li$^+$-FEC coordination was not measured, the data do not establish whether FEC itself enters the first shell or whether its principal effect is indirect.

\begin{figure}[H]
    \centering
    \includegraphics[width=0.75\linewidth]{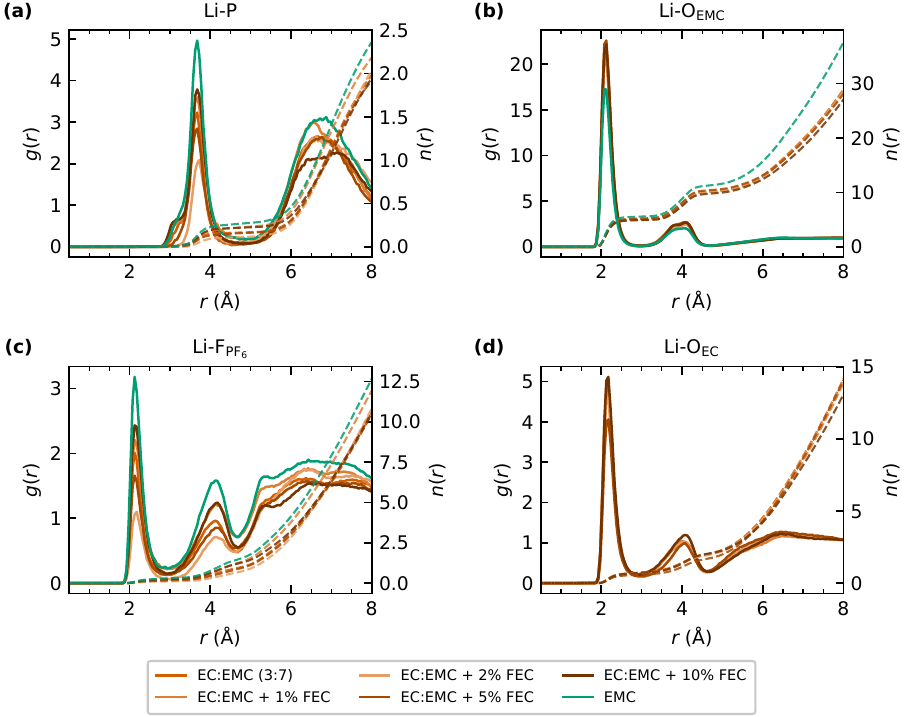}
    \caption{RDF and CN profiles at 233~K for the six selected electrolyte models: (a) Li-P, (b) Li-O$_{\mathrm{EMC}}$, (c) Li-F$_{\mathrm{PF_6}}$ and (d) Li-O$_{\mathrm{EC}}$.}
    \label{fig:rdf_233}
\end{figure}

At 233~K (Figure~\ref{fig:rdf_233}), the six systems exhibit a more differentiated first-shell coordination environment, but cooling does not uniformly strengthen Li$^+$-PF$_6^-$ association. The Li-P and Li-F$_{\mathrm{PF_6}}$ coordination numbers decrease from 298 to 233~K in all six systems, indicating a lower first-shell population of PF$_6^-$ around Li$^+$. For example, in EC:EMC(3:7), the Li-P coordination number decreases from $N=0.252$ to 0.162 and the Li-F$_{\mathrm{PF_6}}$ coordination number from $N=0.299$ to 0.173; corresponding decreases are also observed in neat EMC, from $N=0.317$ to 0.257 and from $N=0.384$ to 0.272, respectively. Neat EMC nevertheless retains the highest absolute Li-P and Li-F$_{\mathrm{PF_6}}$ coordination numbers among the systems examined, indicating a persistently larger PF$_6^-$ first-shell population rather than a low-temperature enhancement of anion coordination. Within the EC:EMC(3:7)+FEC series, the ordering of the anion coordination numbers becomes less systematic at 233~K. The 2~mol\% FEC system exhibits the lowest observed Li-P and Li-F$_{\mathrm{PF_6}}$ coordination numbers ($N=0.108$ and 0.107, respectively), whereas the 10~mol\% FEC composition remains intermediate within the series ($N=0.218$ and 0.241). Because each state point is represented by a single 10~ns production trajectory and molecular mobility is substantially reduced at 233~K, these composition-dependent differences should be interpreted qualitatively rather than as a statistically resolved ranking. In contrast, the Li-O$_{\mathrm{EMC}}$ coordination number increases from 298 to 233~K in all six systems, by approximately $0.13$-$0.45$ coordination units, indicating that the temperature-induced redistribution of the Li$^+$ first coordination shell is accompanied by enhanced EMC-oxygen participation. The 5~mol\% FEC system shows the largest increase ($\Delta N=+0.451$) and the highest Li-O$_{\mathrm{EMC}}$ coordination number within the FEC-containing series at 233~K ($N=5.170$), whereas the 10~mol\% FEC system has the lowest value in this series ($N=4.895$). The Li-O$_{\mathrm{EC}}$ coordination number decreases in all EC-containing systems upon cooling, with the largest decrease observed for 5~mol\% FEC ($N=0.763\rightarrow0.580$), while the decrease at 10~mol\% FEC is more moderate ($N=0.764\rightarrow0.683$). Taken together, these results indicate that cooling redistributes the Li$^+$ first coordination shell rather than uniformly strengthening ion pairing: the first-shell population associated with PF$_6^-$ decreases, whereas EMC-oxygen coordination increases, with 5~mol\% FEC exhibiting the most pronounced changes in the solvent-coordination channels. This interpretation is consistent with the established picture of Li$^+$ solvation by carbonate oxygen atoms and the sensitivity of the local coordination environment to electrolyte composition \cite{Ong2015,Borodin2009,Hou2019,Hou2021}.

The MD simulations also provide a structured composition-temperature conductivity dataset that can be used for rapid electrolyte screening. The motivation is to reduce the computational cost of evaluating nearby formulations: long MD trajectories are expensive, whereas the conductivity response over the sampled composition-temperature domain can be approximated by a supervised surrogate. Because the surrogate is trained directly on MD-derived conductivity labels, its agreement with the atomistic response surface should be interpreted as an internal consistency check rather than as independent validation of the underlying mechanism. The surrogate is therefore treated as an interpolation layer built on top of the atomistic dataset, enabling rapid evaluation of nearby formulations once the mechanistic MD dataset has been established.

\begin{figure}[H]
    \centering
    \includegraphics[width=0.70\textwidth]{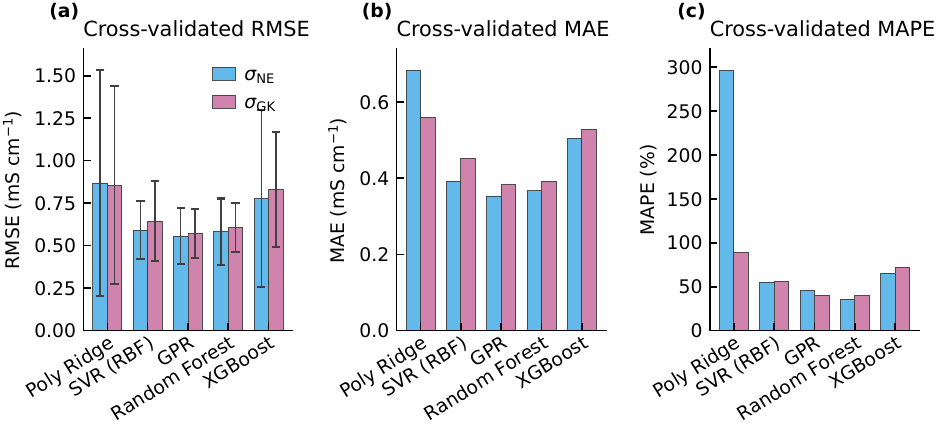}
    \caption{Cross-validated (a) RMSE, (b) MAE and (c) MAPE of the five regression families for the NE ($\sigma_{\mathrm{NE}}$) and GK ($\sigma_{\mathrm{GK}}$) conductivity targets. Error bars in (a) denote the standard deviation across the five outer folds. All four nonlinear families clearly outperform the linear polynomial-ridge baseline; Gaussian process regression is nominally the most accurate for both targets, with random forest and SVR within one fold standard deviation. MAPE is reported only as a diagnostic of the low-conductivity regime.}
    \label{fig:ml_compare}
\end{figure}

Five regression families (polynomial ridge regression, SVR with a radial-basis kernel, GPR, random forest and XGBoost) were benchmarked for each conductivity target under identical group-aware nested cross-validation, in which all temperatures belonging to a given composition were held out together so that every test fold corresponds to unseen electrolyte formulations. The full 154-sample dataset, spanning 11 composition groups and 14 temperatures, was used without any exclusions. The cross-validated errors of all five families are summarized in Table~\ref{tab:ml_models} and Figure~\ref{fig:ml_compare}.

\begin{table}[t]
\centering
\caption{Cross-validated performance of the five regression families for each conductivity target, obtained by group-aware nested cross-validation with composition-disjoint outer folds. Metrics are means over the five composition-disjoint outer folds; RMSE and MAE are in mS~cm$^{-1}$, and the best model for each target is shown in bold. Feature settings and target transformations were selected inside the inner cross-validation; the deployed Gaussian-process models use the $\log(1+y)$ target transform as described in the text.}
\label{tab:ml_models}
\begin{tabular}{lccccccc}
\toprule
 & \multicolumn{3}{c}{$\sigma_{\mathrm{NE}}$} & & \multicolumn{3}{c}{$\sigma_{\mathrm{GK}}$} \\
\cmidrule{2-4}\cmidrule{6-8}
Model & RMSE & MAE & $R^2$ & & RMSE & MAE & $R^2$ \\
\midrule
Polynomial ridge & 0.869 & 0.683 & 0.879 & & 0.857 & 0.560 & 0.860 \\
SVR (RBF)        & 0.592 & 0.393 & 0.962 & & 0.644 & 0.453 & 0.943 \\
Gaussian process & \textbf{0.555} & \textbf{0.352} & \textbf{0.967} & & \textbf{0.572} & \textbf{0.383} & \textbf{0.956} \\
Random forest    & 0.582 & 0.367 & 0.963 & & 0.607 & 0.392 & 0.952 \\
XGBoost          & 0.777 & 0.504 & 0.920 & & 0.830 & 0.528 & 0.906 \\
\bottomrule
\end{tabular}
\end{table}

 All four nonlinear families (SVR, GPR, random forest and XGBoost) outperformed the polynomial-ridge baseline for both targets (Table~\ref{tab:ml_models}), indicating that nonlinear regressors better represent the held-out composition response on this dataset. The most accurate family for both targets was a Gaussian process with a $\log(1+y)$ target transform, reaching RMSE\,$=0.555\pm0.164$~mS~cm$^{-1}$, MAE\,$=0.352$~mS~cm$^{-1}$ and $R^2=0.967$ for $\sigma_{\mathrm{NE}}$, and RMSE\,$=0.572\pm0.145$~mS~cm$^{-1}$, MAE\,$=0.383$~mS~cm$^{-1}$ and $R^2=0.956$ for $\sigma_{\mathrm{GK}}$. In both cases the random forest followed closely (RMSE 0.582 and 0.607) with SVR next (0.592 and 0.644); the leading three nonlinear families are separated by less than one outer-fold standard deviation, so the identity of the single best family should not be over-interpreted; with only five composition-disjoint outer folds, formal significance testing between families is not meaningful, and this comparison is descriptive rather than inferential. Within each target the differences among the three leading nonlinear models are smaller than their fold-to-fold standard deviations ($\pm0.15$-0.20~mS~cm$^{-1}$) and are therefore not statistically distinguishable on a dataset of this size; the practical message is therefore robustness, in the sense that several distinct nonlinear regressors reproduce the conductivity surface to similar accuracy, rather than the primacy of any single algorithm. Because the folds were composition-disjoint, these errors reflect prediction on electrolyte compositions never seen during training, which is substantially more demanding than random sample splitting. Multi-output models that learned the two conductivity targets jointly performed on par with the separate surrogates (best multi-output RMSE 0.553 versus 0.555 for $\sigma_{\mathrm{NE}}$ and 0.571 versus 0.572 for $\sigma_{\mathrm{GK}}$, both within fold variability), so the simpler single-target models were retained for deployment.

A useful reference point for these errors is the reported within-trajectory statistical uncertainty of the training labels. The mean GK sampling standard deviation across the dataset is 0.170~mS~cm$^{-1}$ (median relative uncertainty $\approx9.5\%$). The mean reported GK sampling standard deviation (0.170~mS~cm$^{-1}$) is smaller than the surrogate RMSE (0.572~mS~cm$^{-1}$), indicating that within-trajectory label uncertainty is not the only source of prediction error. A variance decomposition is not attempted because the reported label uncertainties, model residuals and composition-level sampling are not guaranteed to be independent. The MAPE values in Figure~\ref{fig:ml_compare}(c) are reported only as a diagnostic of the low-conductivity regime rather than as a model-selection metric: they are inflated by the vanishingly small conductivities reached at the lowest temperatures, where small absolute errors produce large percentage errors (the $\log(1+y)$ transform partly absorbs this multiplicative structure). Absolute-error metrics (RMSE and MAE) remain the primary measures of surrogate quality here.

\begin{figure}[H]
    \centering
    \includegraphics[width=0.70\textwidth]{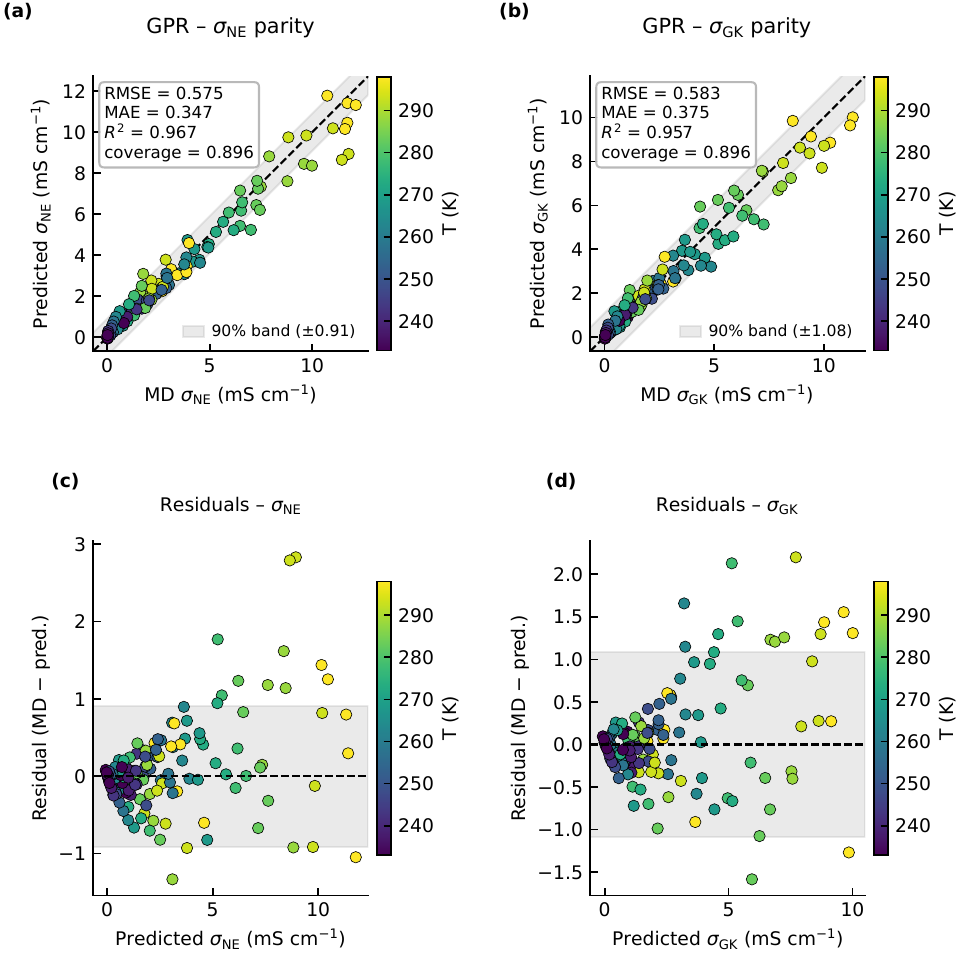}
    \caption{Out-of-fold predictions for the best single-target conductivity surrogates trained on the MD dataset. (a,b) True-versus-predicted parity and (c,d) residual diagnostics for the Gaussian-process models of total NE conductivity ($\sigma_{\mathrm{NE}}$) and total GK conductivity ($\sigma_{\mathrm{GK}}$), both with a $\log(1+y)$ target transform. Each point is a held-out state in composition-disjoint grouped cross-validation, colored by temperature; the shaded band marks the empirical 90\% out-of-fold residual interval} and the inset reports pooled out-of-fold metrics
    \label{fig:ml_parity}
\end{figure}

The out-of-fold parity plots in Figure~\ref{fig:ml_parity}(a,b) show that the selected surrogates reproduce the conductivity trends with only moderate scatter around the diagonal across the full conductivity range (pooled out-of-fold RMSE 0.575 and 0.583~mS~cm$^{-1}$, $R^2=0.967$ and 0.957 for $\sigma_{\mathrm{NE}}$ and $\sigma_{\mathrm{GK}}$, respectively), confirming that the dominant dependence on temperature and composition has been learned. The residual diagnostics in Figure~\ref{fig:ml_parity}(c,d) reveal mild heteroscedasticity: the largest deviations occur at the high-conductivity, high-temperature edge of the dataset, where the response is steepest. Resolved by temperature band, the out-of-fold RMSE decreases from 1.02 (0.83)~mS~cm$^{-1}$ at 288-298~K to 0.11 (0.13)~mS~cm$^{-1}$ at 233-243~K for $\sigma_{\mathrm{NE}}$ ($\sigma_{\mathrm{GK}}$), while the band-wise relative error remains roughly constant at 16-29\%, so the low-temperature regime that motivates this study is predicted no worse, in relative terms, than the ambient one. Empirical 90\% out-of-fold residual bands have half-widths of $\pm0.91$~mS~cm$^{-1}$ ($\sigma_{\mathrm{NE}}$) and $\pm1.08$~mS~cm$^{-1}$ ($\sigma_{\mathrm{GK}}$). Leave-one-outer-fold-out recalibration gives coverages of 0.903 and 0.870, respectively. These values are useful diagnostics of predictive spread, but they are not presented as exact distribution-free conformal coverage because temperatures within a composition form correlated groups.

\begin{figure}[H]
    \centering
    \includegraphics[width=0.70\textwidth]{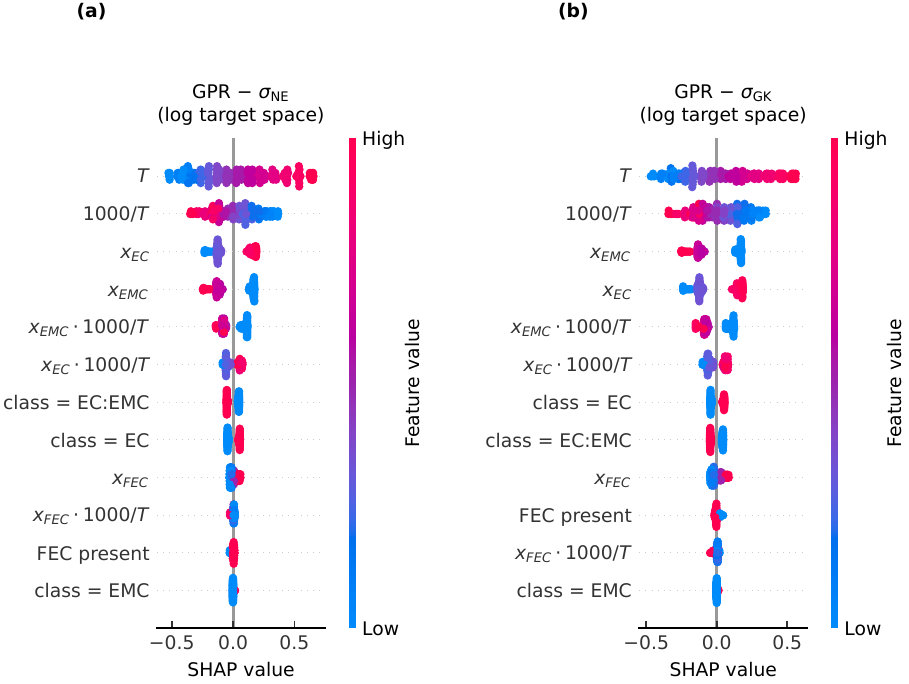}
    \caption{SHAP (Shapley additive explanation) feature attributions \cite{lundberg2017} for the deployed surrogates: Gaussian process regression for (a) $\sigma_{\mathrm{NE}}$ and (b) $\sigma_{\mathrm{GK}}$, both with a $\log(1+y)$ target (attributions shown in the models' log-target space; computed with the model-agnostic permutation explainer). Each point is one electrolyte state; horizontal position gives the signed contribution of a feature to the prediction and color encodes the feature value. Features are ordered by mean absolute attribution.}
    \label{fig:ml_shap}
\end{figure}

To describe which encoded variables the fitted surrogates rely on, SHAP attributions \cite{lundberg2017} were computed for the two deployed models (Figure~\ref{fig:ml_shap}). The deterministic relationship between $T$ and $1000/T$, the unity constraint on the solvent fractions, and the derivation of the interaction terms from their parent variables mean that these attributions do not constitute a unique or causal decomposition of physical importance The two conductivity targets return an essentially common hierarchy. For $\sigma_{\mathrm{NE}}$ the attribution is dominated by temperature ($T$, mean $|$SHAP$|=0.298$ in log-target units, plus $1000/T$ at 0.162; the two rows are redundant encodings of a single physical variable and are read jointly as the temperature attribution), followed by the base-solvent fractions ($x_{\mathrm{EC}}$ at 0.149 and $x_{\mathrm{EMC}}$ at 0.148). The $\sigma_{\mathrm{GK}}$ model reproduces the same ordering ($T$ at 0.254, $1000/T$ at 0.154, $x_{\mathrm{EMC}}$ at 0.149 and $x_{\mathrm{EC}}$ at 0.148), and in both models the leading cross term is the EMC-content cooling interaction $x_{\mathrm{EMC}}\,(1000/T)$ (0.090 for $\sigma_{\mathrm{NE}}$ and 0.094 for $\sigma_{\mathrm{GK}}$). At the descriptive level, this hierarchy mirrors the stronger temperature sensitivity of the EMC-containing MD data; it should not be interpreted as independent evidence for the atomistic mechanism because the surrogate was trained on those same MD outputs The one-hot solvent-family labels carried only minor attribution ($\le0.047$), since the continuous composition fractions already encode most of the solvent identity, although including them gave a small accuracy gain within one cross-validation standard deviation. The FEC descriptors carry the smallest individual attributions ($\le0.037$); given the descriptor dependence noted above, this is interpreted only as evidence that the fitted surrogate relies more strongly on temperature and host-solvent encoding than on the explicit FEC variables

\begin{figure}[H]
    \centering
    \includegraphics[width=0.70\textwidth]{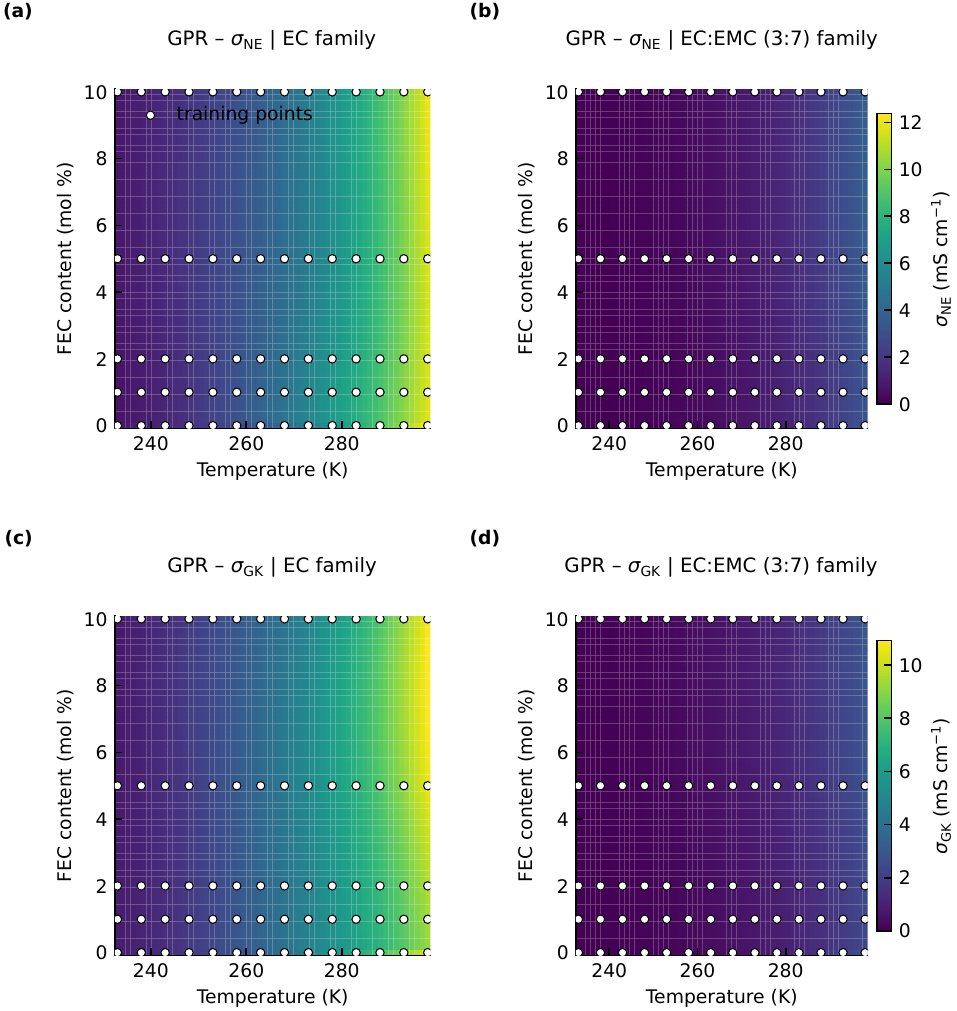}
    \caption{Conductivity maps predicted by the fitted surrogate models over temperature-FEC space for the EC and EC/EMC (3:7, w/w) electrolyte families. The top row shows $\sigma_{\mathrm{NE}}$ and the bottom row shows $\sigma_{\mathrm{GK}}$, both from the selected Gaussian-process surrogates; the left column corresponds to the EC family and the right column to the EC/EMC family. Markers indicate the MD conditions used for training.}
    \label{fig:ml_maps}
\end{figure}

The SHAP analysis identifies temperature and host-solvent composition as the dominant encoded variables in both conductivity surrogates. Their combined effect can be visualized directly in the dense surrogate maps shown in Figure~\ref{fig:ml_maps}, which provide a continuous interpolation of the MD labels: conductivity increases with temperature in all panels, the present model assigns substantially higher cold-regime conductivity to the EC family than to the EC/EMC family, and the high-FEC EC surfaces show reduced retention near the cold edge of the sampled domain. Two limitations should be borne in mind when reading these maps. First, both deployed surrogates are Gaussian processes and therefore produce smooth, continuous surfaces; this smoothness is a property of the kernel interpolant and not evidence that the underlying response is equally smooth, and neither surrogate should be extrapolated beyond the training envelope. The maps must therefore be read strictly as interpolation within the simulated composition-temperature window. Second, the generalization claim is bounded by the modest number of composition groups (11) available for held-out validation. Within these bounds, the practical advantage is decisive: once trained, the surrogate evaluates a composition-temperature point in milliseconds, whereas each of the 154 MD state points on the simulated grid required 15~ns of equilibration and production trajectory per temperature. The surrogate can therefore rank dense grids of intermediate formulations within the sampled composition-temperature domain that would be expensive to cover by direct simulation. In this role, MD supplies the atomistic transport labels and structural analysis while the surrogate supplies rapid interpolation for prioritizing additional simulations; neither layer establishes experimental phase stability or quantitative conductivity without external validation.

The model hierarchy observed here is broadly consistent with the diversity of high-performing regression approaches reported in recent machine-learning studies of electrolyte transport. Shi~\textit{et al.} reported random forest among the top performers for ionic-conductivity prediction in an MD-driven electrolyte screen \cite{shi2025}; Ahmed and Shah found tree ensembles decisively outperforming a neural network on a large experimental ionic-liquid database, attributing this to their robustness on moderately sized tabular data \cite{ahmed2026}; and Chang~\textit{et al.} benchmarked gradient boosting, random forest, and Gaussian process regression for lithium-metal-battery electrolytes and selected the Gaussian process for its superior cross-validated generalization \cite{chang2025}, the same model family that is nominally the most accurate for both conductivity targets here.

The relative weakness of XGBoost in our benchmark also has a transparent origin: under composition-disjoint validation, the folds that hold out boundary compositions of the FEC axis require mild compositional extrapolation, and tree ensembles predict constant plateaus outside their training hull, whereas smooth-kernel interpolants such as GPR and SVR degrade more gracefully there. In the outer fold holding out the FEC-axis boundary compositions, the XGBoost RMSE rises to 1.42~mS~cm$^{-1}$, compared with 0.75~mS~cm$^{-1}$ for the Gaussian process. More generally, the benchmark shows that well-regularized nonlinear regressors—including kernel methods, Gaussian processes, and tree ensembles—can capture the composition and temperature dependence of conductivity while substantially outperforming linear baselines and avoiding the overfitting risks associated with over-parameterized neural networks on data of this scale.

Beyond algorithm selection, the group-aware nested cross-validation adopted here addresses a common vulnerability in data-driven materials science: information leakage between training and test partitions. The composition-disjoint splits ensure that the reported errors reflect generalization to held-out formulations within the sampled design space rather than interpolation between adjacent temperatures of a formulation already present in training \cite{chang2025}. Complementary deep-learning approaches target broader and more discontinuous chemical spaces, for example graph-attention and equivariant graph networks for electrolyte screening \cite{wang2026,sagireddy2026}. For the smooth, continuous EC/EMC/FEC temperature space studied here, however, the lightweight nonlinear surrogates are sufficient and computationally efficient. The attribution pattern is qualitatively similar to feature-importance trends reported for other electrolyte datasets \cite{ahmed2026}, but the chemically distinct data spaces and the correlated descriptors used here preclude treating this similarity as external validation. Overall, the surrogate provides a computationally efficient, composition-disjoint cross-validated interpolation layer for the present MD dataset

The present results should nevertheless be interpreted within several important boundaries. First, the classical charge-scaled nonpolarizable force field does not quantitatively reproduce all experimental transport properties, including the absolute conductivity and room-temperature ranking of the EC and EC/EMC systems. Second, each composition-temperature state is represented by a single 10~ns production trajectory, and the slowest low-temperature states provide limited diffusive sampling; consequently, small differences among FEC loadings are not regarded as statistically resolved. Third, the cooling protocol does not establish equilibrium liquidus boundaries or crystallization kinetics, and the simulated low-temperature states may represent metastable or supercooled liquid-like configurations rather than experimentally stable homogeneous liquids. Fourth, the structural analysis does not include direct Li$^+$-FEC coordination, limiting mechanistic interpretation of FEC-specific first-shell effects. Finally, the machine-learning surrogates are trained on only 11 composition groups and provide interpolation of the MD response surface rather than independent physical validation or experimentally calibrated prediction

\section{Conclusions}

This study used classical MD simulations to examine how temperature, solvent composition, and FEC loading control Li$^+$ transport in LiPF$_6$-based carbonate electrolytes. Within the present force field and cooling protocol, host-solvent composition produces the largest separation in the transport response: EC-based systems have Li$^+$ self-diffusion activation energies of 0.27-0.29~eV, whereas EMC-containing systems give 0.49-0.54~eV and a much stronger reduction of simulated conductivity toward 233~K 

FEC acts as a second-order and nonmonotonic modifier rather than overriding the intrinsic solvent-dependent response. In the EC family, 1-2~mol\% FEC shows a modest tendency toward enhanced Li$^+$ diffusion at intermediate temperatures, whereas 5-10~mol\% FEC has lower conductivity retention at 233~K; the small intermediate-temperature differences are not treated as statistically resolved without independent replicas In EC/EMC mixtures, FEC does not overcome the intrinsic low-temperature transport limitations of the EMC-rich matrix.

The structural analysis provides a complementary molecular interpretation of these trends. The analyzed EC, EMC and PF$_6^-$ coordination channels depend on temperature and composition, with EMC oxygen contributing the largest reported coordination count under the present atom-selection convention Upon cooling to 233~K, the Li$^+$ first coordination shell is redistributed rather than uniformly shifted toward stronger ion pairing: Li-P and Li-F$_{\mathrm{PF_6}}$ coordination decrease, whereas Li-O$_{\mathrm{EMC}}$ coordination increases. The 5~mol\% FEC composition exhibits the most pronounced solvent-coordination changes within the FEC-containing series. Thus, the analyzed coordination channels redistribute rather than showing a universal increase in Li$^+$-PF$_6^-$ association. Direct Li$^+$-FEC coordination was not evaluated, so no mechanistic claim is made about first-shell FEC occupancy

The machine-learning layer provides rapid interpolation of the MD response surface: Gaussian-process models reproduce NE and GK conductivities on composition-disjoint test folds, while the attribution analysis shows that the fitted models rely most strongly on temperature and host-solvent descriptors. Because the surrogates are trained on the MD outputs and the descriptors are correlated, this agreement is an internal model-consistency result rather than independent mechanistic validation

Within the classical charge-scaled nonpolarizable force field, the nominal 1~M LiPF$_6$ EC/EMC/FEC chemistry and the 11 simulated composition groups, EC-based formulations with low FEC loading retain the highest bulk transport in the simulated liquid-like trajectories. This should not be interpreted as a general experimental recommendation at 233~K because the simulations do not establish equilibrium phase stability or crystallization behavior and the force field underpredicts LP57 conductivity and gives a room-temperature solvent ranking that is not quantitatively calibrated to experiment More broadly, the study demonstrates that low-temperature electrolyte optimization requires simultaneous consideration of solvent-controlled mobility, Li$^+$ coordination, ion correlations, and additive-dependent effects. These conclusions concern bulk transport; FEC-induced interfacial stabilization remains a distinct design objective. Extending the framework to broader chemistries, direct Li$^+$-FEC structural descriptors, longer and replicated low-temperature trajectories, explicit phase-stability checks and targeted experimental validation represents a natural next step toward predictive low-temperature electrolyte design.

\begin{suppinfo}
Tabulated equilibrated densities and GK (STACIE) and NE conductivities for all 154 simulated state points with per-point uncertainties and quality-control flags (Table~\ref{tab:si_conductivity}); per-species self-diffusion coefficients, transference numbers and Haven ratios for the same state points (Table~\ref{tab:si_diffusion}); temperature dependence of the equilibrated mass densities (Figure~\ref{fig:si_density}); a machine-readable STACIE summary including the selected spectral-model form, effective sample count and quality-control flag for every state point; the dataset, refit surrogate models and analysis notebook.

\end{suppinfo}

\bibliography{references_last}

\clearpage

\clearpage

% Supporting Information
\setcounter{table}{0}
\renewcommand{\thetable}{S\arabic{table}}
\setcounter{figure}{0}
\renewcommand{\thefigure}{S\arabic{figure}}

\begingroup
\renewcommand{\baselinestretch}{1}\small
\setlength{\tabcolsep}{5pt}
\renewcommand{\arraystretch}{0.57}
\begin{longtable}{lrrll}
\caption{Equilibrated mass densities ($\rho$, g\,cm$^{-3}$) and Green--Kubo (STACIE) and Nernst--Einstein ionic conductivities for all 154 simulated state points (11 electrolytes $\times$ 14 temperatures). The salt is 1~M LiPF$_6$ throughout; EC:EMC (3:7) denotes the 3:7 (w/w) solvent blend and FEC contents are in mol\,\%. Densities are the NPT-equilibrated values at which the NVT production runs were performed. Conductivity uncertainties are one standard deviation. $^{a}$Relative GK uncertainty exceeds 20\,\%. $^{b}$Fewer than the recommended effective number of spectrum points in the STACIE fit (lowest-temperature rows); treat as order-of-magnitude estimates. Both markers are shown where both conditions hold. The 154 rows of this table constitute, unmodified, the training and evaluation dataset of the machine-learning benchmark (Table~\ref{tab:ml_models}).}\label{tab:si_conductivity}\\
\toprule
Electrolyte & $T$ (K) & $\rho$ (g\,cm$^{-3}$) & $\sigma_{\mathrm{GK}}$ (mS\,cm$^{-1}$) & $\sigma_{\mathrm{NE}}$ (mS\,cm$^{-1}$) \\
\midrule
\endfirsthead
\multicolumn{5}{l}{\footnotesize Table~\ref{tab:si_conductivity} (continued)}\\
\toprule
Electrolyte & $T$ (K) & $\rho$ (g\,cm$^{-3}$) & $\sigma_{\mathrm{GK}}$ (mS\,cm$^{-1}$) & $\sigma_{\mathrm{NE}}$ (mS\,cm$^{-1}$) \\
\midrule
\endhead
\bottomrule
\endfoot
EC & 298 & 1.375 & 10.3 $\pm$ 0.5 & 11.7 $\pm$ 1.1 \\
 & 293 & 1.373 & 9.91 $\pm$ 0.55 & 11.8 $\pm$ 1.6 \\
 & 288 & 1.386 & 7.92 $\pm$ 0.45 & 8.78 $\pm$ 1.22 \\
 & 283 & 1.393 & 6.47 $\pm$ 0.38 & 7.28 $\pm$ 0.66 \\
 & 278 & 1.402 & 4.33 $\pm$ 0.50 & 6.47 $\pm$ 0.55 \\
 & 273 & 1.407 & 5.19 $\pm$ 0.34 & 4.95 $\pm$ 0.48 \\
 & 268 & 1.408 & 4.58 $\pm$ 0.33 & 4.50 $\pm$ 0.47 \\
 & 263 & 1.415 & 3.83 $\pm$ 0.36 & 3.48 $\pm$ 0.55 \\
 & 258 & 1.427 & 2.20 $\pm$ 0.17 & 2.48 $\pm$ 0.27 \\
 & 253 & 1.431 & 2.30 $\pm$ 0.19 & 2.14 $\pm$ 0.25 \\
 & 248 & 1.430 & 1.92 $\pm$ 0.16 & 1.97 $\pm$ 0.17 \\
 & 243 & 1.441 & 1.44 $\pm$ 0.13 & 1.34 $\pm$ 0.17 \\
 & 238 & 1.444 & 0.906 $\pm$ 0.100 & 1.01 $\pm$ 0.12 \\
 & 233 & 1.448 & 1.07 $\pm$ 0.12 & 0.764 $\pm$ 0.106 \\
\addlinespace[1.5pt]
EC + 1\% FEC & 298 & 1.371 & 9.41 $\pm$ 0.48 & 11.7 $\pm$ 2.1 \\
 & 293 & 1.385 & 8.14 $\pm$ 0.44 & 9.72 $\pm$ 0.92 \\
 & 288 & 1.384 & 8.08 $\pm$ 0.41 & 9.59 $\pm$ 1.12 \\
 & 283 & 1.397 & 4.36 $\pm$ 0.58 & 7.36 $\pm$ 0.73 \\
 & 278 & 1.394 & 7.25 $\pm$ 0.42 & 6.53 $\pm$ 0.66 \\
 & 273 & 1.394 & 5.50 $\pm$ 0.34 & 6.17 $\pm$ 1.16 \\
 & 268 & 1.407 & 4.12 $\pm$ 0.27 & 4.94 $\pm$ 0.33 \\
 & 263 & 1.404 & 4.86 $\pm$ 0.33 & 4.53 $\pm$ 0.38 \\
 & 258 & 1.419 & 2.50 $\pm$ 0.19 & 2.89 $\pm$ 0.38 \\
 & 253 & 1.427 & 2.69 $\pm$ 0.20 & 2.64 $\pm$ 0.20 \\
 & 248 & 1.436 & 1.56 $\pm$ 0.12 & 1.48 $\pm$ 0.26 \\
 & 243 & 1.443 & 1.16 $\pm$ 0.11 & 1.29 $\pm$ 0.18 \\
 & 238 & 1.446 & 0.916 $\pm$ 0.089 & 0.836 $\pm$ 0.109 \\
 & 233 & 1.449 & 0.936 $\pm$ 0.089 & 0.633 $\pm$ 0.081 \\
\addlinespace[1.5pt]
EC + 2\% FEC & 298 & 1.379 & 8.58 $\pm$ 0.41 & 10.7 $\pm$ 1.2 \\
 & 293 & 1.382 & 8.91 $\pm$ 0.53 & 11.0 $\pm$ 0.8 \\
 & 288 & 1.387 & 7.17 $\pm$ 0.41 & 7.91 $\pm$ 1.39 \\
 & 283 & 1.397 & 5.91 $\pm$ 0.34 & 7.31 $\pm$ 1.19 \\
 & 278 & 1.401 & 5.67 $\pm$ 0.33 & 6.58 $\pm$ 1.06 \\
 & 273 & 1.405 & 4.48 $\pm$ 0.28 & 5.66 $\pm$ 0.64 \\
 & 268 & 1.417 & 3.67 $\pm$ 0.25 & 3.92 $\pm$ 0.66 \\
 & 263 & 1.419 & 3.12 $\pm$ 0.20 & 3.86 $\pm$ 0.42 \\
 & 258 & 1.420 & 3.14 $\pm$ 0.22 & 2.89 $\pm$ 0.48 \\
 & 253 & 1.425 & 2.47 $\pm$ 0.27 & 2.88 $\pm$ 0.45 \\
 & 248 & 1.431 & 1.77 $\pm$ 0.20 & 1.92 $\pm$ 0.17 \\
 & 243 & 1.440 & 1.34 $\pm$ 0.14 & 1.21 $\pm$ 0.15 \\
 & 238 & 1.444 & 0.992 $\pm$ 0.100 & 1.04 $\pm$ 0.13 \\
 & 233 & 1.458 & 0.867 $\pm$ 0.089 & 0.669 $\pm$ 0.069 \\
\addlinespace[1.5pt]
EC + 5\% FEC & 298 & 1.385 & 11.2 $\pm$ 0.5 & 12.1 $\pm$ 1.5 \\
 & 293 & 1.392 & 9.34 $\pm$ 0.53 & 8.83 $\pm$ 1.06 \\
 & 288 & 1.386 & 8.50 $\pm$ 0.56 & 9.98 $\pm$ 1.33 \\
 & 283 & 1.408 & 5.17 $\pm$ 0.31 & 6.48 $\pm$ 0.84 \\
 & 278 & 1.400 & 6.81 $\pm$ 0.42 & 5.94 $\pm$ 1.05 \\
 & 273 & 1.407 & 5.88 $\pm$ 0.34 & 5.31 $\pm$ 0.47 \\
 & 268 & 1.414 & 3.91 $\pm$ 0.34 & 4.25 $\pm$ 0.56 \\
 & 263 & 1.417 & 4.39 $\pm$ 0.27 & 3.77 $\pm$ 0.34 \\
 & 258 & 1.431 & 3.22 $\pm$ 0.22 & 2.92 $\pm$ 0.34 \\
 & 253 & 1.429 & 2.45 $\pm$ 0.19 & 2.66 $\pm$ 0.42 \\
 & 248 & 1.435 & 1.86 $\pm$ 0.16 & 2.06 $\pm$ 0.31 \\
 & 243 & 1.445 & 1.28 $\pm$ 0.11 & 1.44 $\pm$ 0.32 \\
 & 238 & 1.450 & 0.938 $\pm$ 0.097 & 0.981 $\pm$ 0.116 \\
 & 233 & 1.452 & 0.673 $\pm$ 0.131 & 0.816 $\pm$ 0.184 \\
\addlinespace[1.5pt]
EC + 10\% FEC & 298 & 1.388 & 11.3 $\pm$ 0.5 & 11.6 $\pm$ 1.2 \\
 & 293 & 1.389 & 10.0 $\pm$ 0.6 & 11.4 $\pm$ 0.7 \\
 & 288 & 1.410 & 7.22 $\pm$ 0.41 & 7.48 $\pm$ 0.70 \\
 & 283 & 1.408 & 6.08 $\pm$ 0.38 & 7.45 $\pm$ 0.81 \\
 & 278 & 1.408 & 6.28 $\pm$ 0.39 & 7.02 $\pm$ 1.09 \\
 & 273 & 1.421 & 5.11 $\pm$ 0.34 & 4.89 $\pm$ 0.64 \\
 & 268 & 1.427 & 3.53 $\pm$ 0.23 & 4.17 $\pm$ 0.44 \\
 & 263 & 1.429 & 3.61 $\pm$ 0.23 & 3.77 $\pm$ 0.41 \\
 & 258 & 1.433 & 3.23 $\pm$ 0.34 & 3.02 $\pm$ 0.31 \\
 & 253 & 1.434 & 2.41 $\pm$ 0.19 & 2.39 $\pm$ 0.30 \\
 & 248 & 1.443 & 2.16 $\pm$ 0.17 & 1.72 $\pm$ 0.23 \\
 & 243 & 1.449 & 1.64 $\pm$ 0.14 & 1.69 $\pm$ 0.14 \\
 & 238 & 1.463 & 0.747 $\pm$ 0.097 & 0.878 $\pm$ 0.139 \\
 & 233 & 1.467 & 0.627 $\pm$ 0.075 & 0.675 $\pm$ 0.080 \\
\addlinespace[1.5pt]
EC:EMC (3:7) & 298 & 1.205 & 3.22 $\pm$ 0.23 & 3.88 $\pm$ 0.38 \\
 & 293 & 1.223 & 1.83 $\pm$ 0.16 & 2.16 $\pm$ 0.20 \\
 & 288 & 1.223 & 1.89 $\pm$ 0.16 & 1.61 $\pm$ 0.25 \\
 & 283 & 1.228 & 1.43 $\pm$ 0.14 & 1.66 $\pm$ 0.31 \\
 & 278 & 1.231 & 0.628 $\pm$ 0.133$^{a}$ & 1.23 $\pm$ 0.19 \\
 & 273 & 1.244 & 0.720 $\pm$ 0.077 & 0.608 $\pm$ 0.139 \\
 & 268 & 1.246 & 0.481 $\pm$ 0.056 & 0.623 $\pm$ 0.112 \\
 & 263 & 1.248 & 0.431 $\pm$ 0.050 & 0.384 $\pm$ 0.080 \\
 & 258 & 1.251 & 0.160 $\pm$ 0.029 & 0.200 $\pm$ 0.052 \\
 & 253 & 1.262 & 0.100 $\pm$ 0.021$^{a}$ & 0.0527 $\pm$ 0.0184 \\
 & 248 & 1.271 & 0.0578 $\pm$ 0.0152$^{a}$ & 0.0647 $\pm$ 0.0189 \\
 & 243 & 1.272 & 0.0786 $\pm$ 0.0139 & 0.0370 $\pm$ 0.0191 \\
 & 238 & 1.273 & 0.0939 $\pm$ 0.0155$^{b}$ & 0.0241 $\pm$ 0.0170 \\
 & 233 & 1.275 & 0.0491 $\pm$ 0.0105$^{a}$ & 0.0239 $\pm$ 0.0122 \\
\addlinespace[1.5pt]
EC:EMC (3:7) + 1\% FEC & 298 & 1.217 & 2.44 $\pm$ 0.19 & 3.42 $\pm$ 0.48 \\
 & 293 & 1.223 & 1.55 $\pm$ 0.14 & 2.19 $\pm$ 0.27 \\
 & 288 & 1.228 & 1.69 $\pm$ 0.12 & 2.16 $\pm$ 0.20 \\
 & 283 & 1.231 & 1.25 $\pm$ 0.11 & 1.44 $\pm$ 0.13 \\
 & 278 & 1.239 & 0.814 $\pm$ 0.080 & 1.10 $\pm$ 0.23 \\
 & 273 & 1.242 & 0.514 $\pm$ 0.064 & 0.770 $\pm$ 0.136 \\
 & 268 & 1.242 & 0.705 $\pm$ 0.069 & 0.769 $\pm$ 0.119 \\
 & 263 & 1.252 & 0.439 $\pm$ 0.056 & 0.356 $\pm$ 0.102 \\
 & 258 & 1.253 & 0.281 $\pm$ 0.037 & 0.270 $\pm$ 0.058 \\
 & 253 & 1.265 & 0.143 $\pm$ 0.026 & 0.183 $\pm$ 0.056 \\
 & 248 & 1.263 & 0.0748 $\pm$ 0.0155$^{a}$ & 0.127 $\pm$ 0.023 \\
 & 243 & 1.276 & 0.0659 $\pm$ 0.0122 & 0.0484 $\pm$ 0.0163 \\
 & 238 & 1.278 & 0.0333 $\pm$ 0.0100$^{a}$ & 0.0277 $\pm$ 0.0111 \\
 & 233 & 1.287 & 0.0300 $\pm$ 0.0073$^{a}$ & 0.0133 $\pm$ 0.0089 \\
\addlinespace[1.5pt]
EC:EMC (3:7) + 2\% FEC & 298 & 1.219 & 2.58 $\pm$ 0.19 & 3.12 $\pm$ 0.27 \\
 & 293 & 1.226 & 1.83 $\pm$ 0.16 & 2.52 $\pm$ 0.31 \\
 & 288 & 1.231 & 1.38 $\pm$ 0.12 & 1.80 $\pm$ 0.17 \\
 & 283 & 1.240 & 1.32 $\pm$ 0.11 & 1.25 $\pm$ 0.15 \\
 & 278 & 1.240 & 1.00 $\pm$ 0.10 & 1.11 $\pm$ 0.15 \\
 & 273 & 1.243 & 0.642 $\pm$ 0.077 & 1.00 $\pm$ 0.19 \\
 & 268 & 1.254 & 0.406 $\pm$ 0.055 & 0.517 $\pm$ 0.059 \\
 & 263 & 1.256 & 0.344 $\pm$ 0.048 & 0.459 $\pm$ 0.081 \\
 & 258 & 1.268 & 0.158 $\pm$ 0.027 & 0.177 $\pm$ 0.028 \\
 & 253 & 1.269 & 0.110 $\pm$ 0.022 & 0.131 $\pm$ 0.041 \\
 & 248 & 1.267 & 0.130 $\pm$ 0.022 & 0.125 $\pm$ 0.024 \\
 & 243 & 1.277 & 0.0552 $\pm$ 0.0133$^{a,b}$ & 0.0530 $\pm$ 0.0187 \\
 & 238 & 1.279 & 0.0495 $\pm$ 0.0111$^{a}$ & 0.0428 $\pm$ 0.0241 \\
 & 233 & 1.280 & 0.0170 $\pm$ 0.0056$^{a,b}$ & 0.0195 $\pm$ 0.0172 \\
\addlinespace[1.5pt]
EC:EMC (3:7) + 5\% FEC & 298 & 1.216 & 3.12 $\pm$ 0.22 & 3.86 $\pm$ 0.45 \\
 & 293 & 1.233 & 1.81 $\pm$ 0.14 & 2.72 $\pm$ 0.31 \\
 & 288 & 1.237 & 1.67 $\pm$ 0.19 & 2.03 $\pm$ 0.38 \\
 & 283 & 1.237 & 1.55 $\pm$ 0.15 & 1.86 $\pm$ 0.23 \\
 & 278 & 1.250 & 0.894 $\pm$ 0.084 & 1.31 $\pm$ 0.12 \\
 & 273 & 1.250 & 0.948 $\pm$ 0.089 & 0.988 $\pm$ 0.100 \\
 & 268 & 1.253 & 0.678 $\pm$ 0.070 & 0.613 $\pm$ 0.073 \\
 & 263 & 1.261 & 0.537 $\pm$ 0.062 & 0.503 $\pm$ 0.056 \\
 & 258 & 1.269 & 0.244 $\pm$ 0.036 & 0.283 $\pm$ 0.062 \\
 & 253 & 1.273 & 0.230 $\pm$ 0.028 & 0.172 $\pm$ 0.055 \\
 & 248 & 1.279 & 0.158 $\pm$ 0.041$^{a}$ & 0.150 $\pm$ 0.031 \\
 & 243 & 1.292 & 0.0238 $\pm$ 0.0089$^{a}$ & 0.0505 $\pm$ 0.0184 \\
 & 238 & 1.291 & 0.0350 $\pm$ 0.0084$^{a}$ & 0.0236 $\pm$ 0.0202 \\
 & 233 & 1.295 & 0.0339 $\pm$ 0.0081$^{a}$ & 0.0280 $\pm$ 0.0087 \\
\addlinespace[1.5pt]
EC:EMC (3:7) + 10\% FEC & 298 & 1.237 & 2.75 $\pm$ 0.20 & 3.98 $\pm$ 0.72 \\
 & 293 & 1.243 & 2.64 $\pm$ 0.19 & 2.86 $\pm$ 0.16 \\
 & 288 & 1.252 & 1.95 $\pm$ 0.16 & 1.77 $\pm$ 0.33 \\
 & 283 & 1.263 & 1.15 $\pm$ 0.11 & 1.69 $\pm$ 0.16 \\
 & 278 & 1.262 & 1.07 $\pm$ 0.10 & 1.32 $\pm$ 0.18 \\
 & 273 & 1.262 & 0.916 $\pm$ 0.089 & 1.07 $\pm$ 0.15 \\
 & 268 & 1.271 & 0.452 $\pm$ 0.061 & 0.609 $\pm$ 0.083 \\
 & 263 & 1.277 & 0.588 $\pm$ 0.056 & 0.428 $\pm$ 0.089 \\
 & 258 & 1.274 & 0.494 $\pm$ 0.056 & 0.320 $\pm$ 0.097 \\
 & 253 & 1.283 & 0.170 $\pm$ 0.030 & 0.242 $\pm$ 0.037 \\
 & 248 & 1.295 & 0.154 $\pm$ 0.029 & 0.198 $\pm$ 0.050 \\
 & 243 & 1.306 & 0.0859 $\pm$ 0.0153 & 0.0544 $\pm$ 0.0433 \\
 & 238 & 1.306 & 0.0648 $\pm$ 0.0128 & 0.0478 $\pm$ 0.0141 \\
 & 233 & 1.305 & 0.0411 $\pm$ 0.0089$^{a}$ & 0.0233 $\pm$ 0.0102 \\
\addlinespace[1.5pt]
EMC & 298 & 1.140 & 1.77 $\pm$ 0.14 & 2.67 $\pm$ 0.42 \\
 & 293 & 1.157 & 1.05 $\pm$ 0.10 & 1.28 $\pm$ 0.12 \\
 & 288 & 1.163 & 1.08 $\pm$ 0.10 & 1.42 $\pm$ 0.13 \\
 & 283 & 1.163 & 0.747 $\pm$ 0.072 & 1.00 $\pm$ 0.15 \\
 & 278 & 1.165 & 0.472 $\pm$ 0.053 & 0.713 $\pm$ 0.111 \\
 & 273 & 1.181 & 0.334 $\pm$ 0.044 & 0.419 $\pm$ 0.037 \\
 & 268 & 1.180 & 0.228 $\pm$ 0.036 & 0.330 $\pm$ 0.064 \\
 & 263 & 1.192 & 0.105 $\pm$ 0.017 & 0.0858 $\pm$ 0.0163 \\
 & 258 & 1.192 & 0.0811 $\pm$ 0.0142 & 0.0611 $\pm$ 0.0148 \\
 & 253 & 1.198 & 0.0644 $\pm$ 0.0127 & 0.0617 $\pm$ 0.0170 \\
 & 248 & 1.201 & 0.0600 $\pm$ 0.0117 & 0.0511 $\pm$ 0.0117 \\
 & 243 & 1.208 & 0.0378 $\pm$ 0.0158$^{a}$ & 0.0411 $\pm$ 0.0114 \\
 & 238 & 1.215 & 0.0317 $\pm$ 0.0078$^{a}$ & 0.0147 $\pm$ 0.0112 \\
 & 233 & 1.216 & 0.0163 $\pm$ 0.0047$^{a,b}$ & 0.0173 $\pm$ 0.0094 \\
\end{longtable}
\endgroup

\clearpage
% Auto-generated from all_results_en_gk.csv - per-species diffusion, t_Li, Haven ratio
\begingroup
\renewcommand{\baselinestretch}{1}\small
\setlength{\tabcolsep}{4.5pt}
\renewcommand{\arraystretch}{0.65}
\begin{longtable}{lccccc}
\caption{Li$^+$ and PF$_6^-$ self-diffusion coefficients (units of $10^{-7}$~cm$^2$\,s$^{-1}$, one-standard-deviation fit uncertainties), apparent Li$^+$ transference numbers $t_{\mathrm{Li}}$ and Haven ratios $H_R=\sigma_{\mathrm{NE}}/\sigma_{\mathrm{GK}}$ for all 154 simulated state points. Values are computed from the same trajectories as Table~\ref{tab:si_conductivity}.}\label{tab:si_diffusion}\\
\toprule
Electrolyte & $T$ (K) & $D_{\mathrm{Li}}$ & $D_{\mathrm{PF_6}}$ & $t_{\mathrm{Li}}$ & $H_R$ \\
\midrule
\endfirsthead
\multicolumn{6}{l}{Table~\ref{tab:si_diffusion} (continued)}\\
\toprule
Electrolyte & $T$ (K) & $D_{\mathrm{Li}}$ & $D_{\mathrm{PF_6}}$ & $t_{\mathrm{Li}}$ & $H_R$ \\
\midrule
\endhead
\bottomrule
\endfoot
EC & 298 & 13.42 $\pm$ 1.84 & 16.65 $\pm$ 2.28 & 0.446 $\pm$ 0.048 & 1.14 \\
 & 293 & 14.81 $\pm$ 1.68 & 14.93 $\pm$ 3.79 & 0.498 $\pm$ 0.070 & 1.19 \\
 & 288 & 12.19 $\pm$ 1.32 & 9.44 $\pm$ 2.70 & 0.564 $\pm$ 0.075 & 1.11 \\
 & 283 & 8.48 $\pm$ 1.23 & 9.02 $\pm$ 1.00 & 0.485 $\pm$ 0.046 & 1.12 \\
 & 278 & 8.28 $\pm$ 1.17 & 6.93 $\pm$ 0.51 & 0.545 $\pm$ 0.040 & 1.49 \\
 & 273 & 5.76 $\pm$ 0.77 & 5.63 $\pm$ 0.78 & 0.505 $\pm$ 0.048 & 0.95 \\
 & 268 & 5.04 $\pm$ 0.76 & 5.10 $\pm$ 0.76 & 0.497 $\pm$ 0.053 & 0.98 \\
 & 263 & 3.84 $\pm$ 0.53 & 3.84 $\pm$ 1.08 & 0.500 $\pm$ 0.078 & 0.91 \\
 & 258 & 2.71 $\pm$ 0.28 & 2.61 $\pm$ 0.52 & 0.510 $\pm$ 0.056 & 1.13 \\
 & 253 & 2.10 $\pm$ 0.29 & 2.39 $\pm$ 0.42 & 0.468 $\pm$ 0.056 & 0.93 \\
 & 248 & 1.97 $\pm$ 0.10 & 2.07 $\pm$ 0.34 & 0.487 $\pm$ 0.043 & 1.03 \\
 & 243 & 1.35 $\pm$ 0.21 & 1.33 $\pm$ 0.27 & 0.505 $\pm$ 0.064 & 0.93 \\
 & 238 & 0.93 $\pm$ 0.17 & 1.04 $\pm$ 0.16 & 0.472 $\pm$ 0.060 & 1.11 \\
 & 233 & 0.72 $\pm$ 0.10 & 0.73 $\pm$ 0.18 & 0.496 $\pm$ 0.070 & 0.72 \\
\addlinespace[1.5pt]
EC + 1\% FEC & 298 & 14.22 $\pm$ 2.65 & 17.09 $\pm$ 5.06 & 0.454 $\pm$ 0.087 & 1.24 \\
 & 293 & 11.95 $\pm$ 1.24 & 13.33 $\pm$ 2.05 & 0.473 $\pm$ 0.046 & 1.19 \\
 & 288 & 12.86 $\pm$ 2.56 & 11.70 $\pm$ 1.33 & 0.524 $\pm$ 0.057 & 1.19 \\
 & 283 & 9.45 $\pm$ 1.39 & 8.88 $\pm$ 1.18 & 0.516 $\pm$ 0.049 & 1.69 \\
 & 278 & 8.65 $\pm$ 1.53 & 7.38 $\pm$ 0.45 & 0.540 $\pm$ 0.046 & 0.90 \\
 & 273 & 7.46 $\pm$ 1.28 & 7.39 $\pm$ 2.49 & 0.502 $\pm$ 0.095 & 1.12 \\
 & 268 & 5.81 $\pm$ 0.29 & 5.75 $\pm$ 0.72 & 0.503 $\pm$ 0.034 & 1.20 \\
 & 263 & 5.28 $\pm$ 0.75 & 5.17 $\pm$ 0.41 & 0.505 $\pm$ 0.041 & 0.93 \\
 & 258 & 3.29 $\pm$ 0.55 & 3.19 $\pm$ 0.65 & 0.507 $\pm$ 0.066 & 1.16 \\
 & 253 & 3.04 $\pm$ 0.34 & 2.70 $\pm$ 0.29 & 0.529 $\pm$ 0.039 & 0.98 \\
 & 248 & 1.60 $\pm$ 0.46 & 1.54 $\pm$ 0.31 & 0.509 $\pm$ 0.088 & 0.94 \\
 & 243 & 1.48 $\pm$ 0.33 & 1.20 $\pm$ 0.19 & 0.553 $\pm$ 0.067 & 1.11 \\
 & 238 & 0.93 $\pm$ 0.21 & 0.76 $\pm$ 0.08 & 0.552 $\pm$ 0.061 & 0.91 \\
 & 233 & 0.61 $\pm$ 0.12 & 0.64 $\pm$ 0.10 & 0.489 $\pm$ 0.064 & 0.68 \\
\addlinespace[1.5pt]
EC + 2\% FEC & 298 & 12.54 $\pm$ 1.76 & 16.04 $\pm$ 2.74 & 0.439 $\pm$ 0.054 & 1.25 \\
 & 293 & 15.87 $\pm$ 1.67 & 12.89 $\pm$ 1.38 & 0.552 $\pm$ 0.037 & 1.24 \\
 & 288 & 9.93 $\pm$ 1.61 & 10.29 $\pm$ 3.17 & 0.491 $\pm$ 0.087 & 1.10 \\
 & 283 & 8.94 $\pm$ 2.36 & 9.32 $\pm$ 1.80 & 0.490 $\pm$ 0.082 & 1.24 \\
 & 278 & 8.23 $\pm$ 2.55 & 7.88 $\pm$ 0.52 & 0.511 $\pm$ 0.079 & 1.16 \\
 & 273 & 7.26 $\pm$ 1.49 & 6.28 $\pm$ 0.31 & 0.536 $\pm$ 0.053 & 1.26 \\
 & 268 & 4.49 $\pm$ 1.23 & 4.64 $\pm$ 0.91 & 0.492 $\pm$ 0.084 & 1.07 \\
 & 263 & 4.21 $\pm$ 0.68 & 4.61 $\pm$ 0.69 & 0.478 $\pm$ 0.055 & 1.23 \\
 & 258 & 3.30 $\pm$ 0.95 & 3.17 $\pm$ 0.55 & 0.510 $\pm$ 0.084 & 0.92 \\
 & 253 & 3.05 $\pm$ 0.32 & 3.25 $\pm$ 0.92 & 0.484 $\pm$ 0.075 & 1.16 \\
 & 248 & 1.95 $\pm$ 0.25 & 2.17 $\pm$ 0.27 & 0.472 $\pm$ 0.045 & 1.10 \\
 & 243 & 1.27 $\pm$ 0.26 & 1.25 $\pm$ 0.15 & 0.503 $\pm$ 0.060 & 0.90 \\
 & 238 & 1.08 $\pm$ 0.08 & 1.04 $\pm$ 0.24 & 0.509 $\pm$ 0.061 & 1.05 \\
 & 233 & 0.75 $\pm$ 0.10 & 0.57 $\pm$ 0.09 & 0.567 $\pm$ 0.050 & 0.77 \\
\addlinespace[1.5pt]
EC + 5\% FEC & 298 & 16.11 $\pm$ 3.21 & 16.21 $\pm$ 2.33 & 0.498 $\pm$ 0.061 & 1.08 \\
 & 293 & 11.52 $\pm$ 2.38 & 11.52 $\pm$ 1.41 & 0.500 $\pm$ 0.060 & 0.94 \\
 & 288 & 13.42 $\pm$ 2.74 & 12.29 $\pm$ 2.04 & 0.522 $\pm$ 0.066 & 1.17 \\
 & 283 & 8.44 $\pm$ 0.89 & 7.70 $\pm$ 1.92 & 0.523 $\pm$ 0.068 & 1.25 \\
 & 278 & 7.31 $\pm$ 2.00 & 7.29 $\pm$ 1.63 & 0.501 $\pm$ 0.088 & 0.87 \\
 & 273 & 6.28 $\pm$ 0.81 & 6.48 $\pm$ 0.78 & 0.492 $\pm$ 0.044 & 0.90 \\
 & 268 & 4.97 $\pm$ 0.85 & 5.02 $\pm$ 1.01 & 0.498 $\pm$ 0.066 & 1.09 \\
 & 263 & 4.21 $\pm$ 0.61 & 4.44 $\pm$ 0.51 & 0.486 $\pm$ 0.046 & 0.85 \\
 & 258 & 3.00 $\pm$ 0.67 & 3.54 $\pm$ 0.35 & 0.459 $\pm$ 0.061 & 0.91 \\
 & 253 & 2.85 $\pm$ 0.72 & 2.97 $\pm$ 0.60 & 0.489 $\pm$ 0.081 & 1.08 \\
 & 248 & 2.47 $\pm$ 0.51 & 1.94 $\pm$ 0.41 & 0.560 $\pm$ 0.073 & 1.11 \\
 & 243 & 1.46 $\pm$ 0.63 & 1.53 $\pm$ 0.21 & 0.488 $\pm$ 0.114 & 1.12 \\
 & 238 & 0.93 $\pm$ 0.15 & 1.06 $\pm$ 0.18 & 0.468 $\pm$ 0.059 & 1.05 \\
 & 233 & 0.71 $\pm$ 0.23 & 0.91 $\pm$ 0.29 & 0.438 $\pm$ 0.110 & 1.21 \\
\addlinespace[1.5pt]
EC + 10\% FEC & 298 & 15.85 $\pm$ 1.97 & 15.27 $\pm$ 2.58 & 0.509 $\pm$ 0.052 & 1.02 \\
 & 293 & 16.25 $\pm$ 1.58 & 13.94 $\pm$ 1.14 & 0.538 $\pm$ 0.032 & 1.14 \\
 & 288 & 9.10 $\pm$ 0.71 & 10.03 $\pm$ 1.65 & 0.476 $\pm$ 0.046 & 1.04 \\
 & 283 & 9.23 $\pm$ 1.18 & 9.48 $\pm$ 1.67 & 0.493 $\pm$ 0.054 & 1.22 \\
 & 278 & 8.44 $\pm$ 2.26 & 8.86 $\pm$ 1.46 & 0.488 $\pm$ 0.079 & 1.12 \\
 & 273 & 5.27 $\pm$ 1.29 & 6.47 $\pm$ 0.82 & 0.449 $\pm$ 0.068 & 0.96 \\
 & 268 & 5.19 $\pm$ 0.99 & 4.62 $\pm$ 0.33 & 0.529 $\pm$ 0.051 & 1.18 \\
 & 263 & 4.10 $\pm$ 0.51 & 4.58 $\pm$ 0.79 & 0.473 $\pm$ 0.053 & 1.04 \\
 & 258 & 3.34 $\pm$ 0.58 & 3.46 $\pm$ 0.38 & 0.491 $\pm$ 0.051 & 0.93 \\
 & 253 & 2.78 $\pm$ 0.33 & 2.48 $\pm$ 0.55 & 0.529 $\pm$ 0.063 & 0.99 \\
 & 248 & 1.83 $\pm$ 0.27 & 1.85 $\pm$ 0.43 & 0.498 $\pm$ 0.069 & 0.79 \\
 & 243 & 1.63 $\pm$ 0.14 & 1.91 $\pm$ 0.26 & 0.462 $\pm$ 0.040 & 1.03 \\
 & 238 & 0.86 $\pm$ 0.12 & 0.93 $\pm$ 0.26 & 0.480 $\pm$ 0.077 & 1.18 \\
 & 233 & 0.68 $\pm$ 0.09 & 0.66 $\pm$ 0.13 & 0.506 $\pm$ 0.059 & 1.08 \\
\addlinespace[1.5pt]
EC:EMC (3:7) & 298 & 3.77 $\pm$ 0.38 & 6.44 $\pm$ 0.91 & 0.369 $\pm$ 0.040 & 1.20 \\
 & 293 & 2.15 $\pm$ 0.47 & 3.34 $\pm$ 0.26 & 0.392 $\pm$ 0.055 & 1.18 \\
 & 288 & 1.35 $\pm$ 0.10 & 2.68 $\pm$ 0.61 & 0.335 $\pm$ 0.053 & 0.85 \\
 & 283 & 1.44 $\pm$ 0.16 & 2.64 $\pm$ 0.76 & 0.354 $\pm$ 0.070 & 1.16 \\
 & 278 & 1.18 $\pm$ 0.25 & 1.78 $\pm$ 0.38 & 0.399 $\pm$ 0.073 & 1.97 \\
 & 273 & 0.50 $\pm$ 0.14 & 0.92 $\pm$ 0.29 & 0.354 $\pm$ 0.097 & 0.84 \\
 & 268 & 0.45 $\pm$ 0.14 & 0.98 $\pm$ 0.22 & 0.313 $\pm$ 0.083 & 1.30 \\
 & 263 & 0.29 $\pm$ 0.10 & 0.57 $\pm$ 0.15 & 0.334 $\pm$ 0.095 & 0.89 \\
 & 258 & 0.18 $\pm$ 0.02 & 0.26 $\pm$ 0.11 & 0.399 $\pm$ 0.106 & 1.25 \\
 & 253 & 0.05 $\pm$ 0.02 & 0.06 $\pm$ 0.03 & 0.473 $\pm$ 0.173 & 0.52 \\
 & 248 & 0.04 $\pm$ 0.02 & 0.09 $\pm$ 0.04 & 0.321 $\pm$ 0.119 & 1.12 \\
 & 243 & 0.05 $\pm$ 0.03 & 0.03 $\pm$ 0.03 & 0.652 $\pm$ 0.278 & 0.47 \\
 & 238 & 0.02 $\pm$ 0.02 & 0.03 $\pm$ 0.03 & 0.431 $\pm$ 0.329 & 0.26 \\
 & 233 & 0.03 $\pm$ 0.01 & 0.02 $\pm$ 0.02 & 0.652 $\pm$ 0.291 & 0.49 \\
\addlinespace[1.5pt]
EC:EMC (3:7) + 1\% FEC & 298 & 3.00 $\pm$ 1.13 & 5.91 $\pm$ 0.55 & 0.337 $\pm$ 0.087 & 1.40 \\
 & 293 & 2.12 $\pm$ 0.35 & 3.43 $\pm$ 0.57 & 0.382 $\pm$ 0.055 & 1.40 \\
 & 288 & 1.75 $\pm$ 0.15 & 3.63 $\pm$ 0.50 & 0.326 $\pm$ 0.035 & 1.28 \\
 & 283 & 1.27 $\pm$ 0.24 & 2.24 $\pm$ 0.23 & 0.363 $\pm$ 0.049 & 1.15 \\
 & 278 & 0.92 $\pm$ 0.11 & 1.72 $\pm$ 0.54 & 0.347 $\pm$ 0.075 & 1.36 \\
 & 273 & 0.67 $\pm$ 0.24 & 1.14 $\pm$ 0.21 & 0.370 $\pm$ 0.094 & 1.50 \\
 & 268 & 0.69 $\pm$ 0.12 & 1.08 $\pm$ 0.24 & 0.389 $\pm$ 0.069 & 1.09 \\
 & 263 & 0.24 $\pm$ 0.04 & 0.56 $\pm$ 0.22 & 0.297 $\pm$ 0.091 & 0.81 \\
 & 258 & 0.18 $\pm$ 0.03 & 0.41 $\pm$ 0.12 & 0.310 $\pm$ 0.072 & 0.96 \\
 & 253 & 0.14 $\pm$ 0.05 & 0.25 $\pm$ 0.11 & 0.357 $\pm$ 0.127 & 1.29 \\
 & 248 & 0.10 $\pm$ 0.03 & 0.16 $\pm$ 0.04 & 0.380 $\pm$ 0.085 & 1.70 \\
 & 243 & 0.05 $\pm$ 0.02 & 0.05 $\pm$ 0.03 & 0.486 $\pm$ 0.166 & 0.73 \\
 & 238 & 0.02 $\pm$ 0.00 & 0.03 $\pm$ 0.02 & 0.383 $\pm$ 0.159 & 0.83 \\
 & 233 & 0.01 $\pm$ 0.01 & 0.02 $\pm$ 0.01 & 0.344 $\pm$ 0.341 & 0.44 \\
\addlinespace[1.5pt]
EC:EMC (3:7) + 2\% FEC & 298 & 2.88 $\pm$ 0.44 & 5.25 $\pm$ 0.52 & 0.354 $\pm$ 0.042 & 1.21 \\
 & 293 & 2.46 $\pm$ 0.34 & 3.94 $\pm$ 0.71 & 0.384 $\pm$ 0.054 & 1.38 \\
 & 288 & 1.69 $\pm$ 0.19 & 2.78 $\pm$ 0.38 & 0.378 $\pm$ 0.042 & 1.30 \\
 & 283 & 1.25 $\pm$ 0.24 & 1.80 $\pm$ 0.29 & 0.410 $\pm$ 0.060 & 0.95 \\
 & 278 & 0.85 $\pm$ 0.22 & 1.81 $\pm$ 0.27 & 0.318 $\pm$ 0.066 & 1.11 \\
 & 273 & 0.80 $\pm$ 0.19 & 1.55 $\pm$ 0.40 & 0.340 $\pm$ 0.079 & 1.56 \\
 & 268 & 0.41 $\pm$ 0.11 & 0.77 $\pm$ 0.08 & 0.344 $\pm$ 0.066 & 1.27 \\
 & 263 & 0.33 $\pm$ 0.08 & 0.69 $\pm$ 0.17 & 0.325 $\pm$ 0.072 & 1.34 \\
 & 258 & 0.15 $\pm$ 0.04 & 0.23 $\pm$ 0.04 & 0.389 $\pm$ 0.080 & 1.13 \\
 & 253 & 0.11 $\pm$ 0.04 & 0.17 $\pm$ 0.08 & 0.398 $\pm$ 0.136 & 1.20 \\
 & 248 & 0.09 $\pm$ 0.03 & 0.17 $\pm$ 0.04 & 0.358 $\pm$ 0.090 & 0.96 \\
 & 243 & 0.03 $\pm$ 0.01 & 0.08 $\pm$ 0.04 & 0.262 $\pm$ 0.101 & 0.96 \\
 & 238 & 0.04 $\pm$ 0.03 & 0.05 $\pm$ 0.04 & 0.415 $\pm$ 0.277 & 0.86 \\
 & 233 & 0.01 $\pm$ 0.01 & 0.03 $\pm$ 0.03 & 0.309 $\pm$ 0.317 & 1.14 \\
\addlinespace[1.5pt]
EC:EMC (3:7) + 5\% FEC & 298 & 3.79 $\pm$ 0.65 & 6.31 $\pm$ 0.99 & 0.375 $\pm$ 0.054 & 1.24 \\
 & 293 & 2.53 $\pm$ 0.55 & 4.39 $\pm$ 0.58 & 0.365 $\pm$ 0.059 & 1.51 \\
 & 288 & 2.16 $\pm$ 0.27 & 2.90 $\pm$ 0.89 & 0.427 $\pm$ 0.081 & 1.22 \\
 & 283 & 1.85 $\pm$ 0.25 & 2.69 $\pm$ 0.51 & 0.407 $\pm$ 0.057 & 1.20 \\
 & 278 & 1.29 $\pm$ 0.20 & 1.83 $\pm$ 0.20 & 0.413 $\pm$ 0.047 & 1.47 \\
 & 273 & 0.82 $\pm$ 0.18 & 1.49 $\pm$ 0.15 & 0.354 $\pm$ 0.055 & 1.04 \\
 & 268 & 0.54 $\pm$ 0.06 & 0.86 $\pm$ 0.16 & 0.385 $\pm$ 0.050 & 0.90 \\
 & 263 & 0.34 $\pm$ 0.09 & 0.78 $\pm$ 0.09 & 0.306 $\pm$ 0.063 & 0.93 \\
 & 258 & 0.26 $\pm$ 0.07 & 0.35 $\pm$ 0.11 & 0.429 $\pm$ 0.104 & 1.16 \\
 & 253 & 0.12 $\pm$ 0.03 & 0.25 $\pm$ 0.11 & 0.317 $\pm$ 0.116 & 0.75 \\
 & 248 & 0.12 $\pm$ 0.04 & 0.19 $\pm$ 0.05 & 0.384 $\pm$ 0.102 & 0.95 \\
 & 243 & 0.04 $\pm$ 0.01 & 0.06 $\pm$ 0.04 & 0.361 $\pm$ 0.141 & 2.12 \\
 & 238 & 0.02 $\pm$ 0.01 & 0.03 $\pm$ 0.04 & 0.398 $\pm$ 0.348 & 0.67 \\
 & 233 & 0.01 $\pm$ 0.01 & 0.04 $\pm$ 0.02 & 0.170 $\pm$ 0.121 & 0.83 \\
\addlinespace[1.5pt]
EC:EMC (3:7) + 10\% FEC & 298 & 3.60 $\pm$ 0.58 & 6.70 $\pm$ 1.77 & 0.349 $\pm$ 0.070 & 1.45 \\
 & 293 & 2.69 $\pm$ 0.22 & 4.52 $\pm$ 0.31 & 0.373 $\pm$ 0.025 & 1.08 \\
 & 288 & 1.77 $\pm$ 0.19 & 2.56 $\pm$ 0.77 & 0.409 $\pm$ 0.077 & 0.90 \\
 & 283 & 1.70 $\pm$ 0.15 & 2.35 $\pm$ 0.34 & 0.419 $\pm$ 0.041 & 1.47 \\
 & 278 & 1.31 $\pm$ 0.07 & 1.79 $\pm$ 0.41 & 0.423 $\pm$ 0.057 & 1.23 \\
 & 273 & 0.77 $\pm$ 0.24 & 1.70 $\pm$ 0.25 & 0.310 $\pm$ 0.075 & 1.17 \\
 & 268 & 0.49 $\pm$ 0.08 & 0.88 $\pm$ 0.17 & 0.358 $\pm$ 0.058 & 1.35 \\
 & 263 & 0.31 $\pm$ 0.08 & 0.64 $\pm$ 0.18 & 0.324 $\pm$ 0.083 & 0.73 \\
 & 258 & 0.25 $\pm$ 0.05 & 0.44 $\pm$ 0.20 & 0.367 $\pm$ 0.118 & 0.65 \\
 & 253 & 0.17 $\pm$ 0.05 & 0.35 $\pm$ 0.06 & 0.322 $\pm$ 0.080 & 1.42 \\
 & 248 & 0.08 $\pm$ 0.02 & 0.33 $\pm$ 0.10 & 0.184 $\pm$ 0.061 & 1.29 \\
 & 243 & 0.05 $\pm$ 0.03 & 0.06 $\pm$ 0.08 & 0.430 $\pm$ 0.356 & 0.63 \\
 & 238 & 0.03 $\pm$ 0.01 & 0.06 $\pm$ 0.03 & 0.310 $\pm$ 0.114 & 0.74 \\
 & 233 & 0.02 $\pm$ 0.01 & 0.02 $\pm$ 0.02 & 0.507 $\pm$ 0.220 & 0.56 \\
\addlinespace[1.5pt]
EMC & 298 & 3.28 $\pm$ 0.64 & 3.97 $\pm$ 0.95 & 0.452 $\pm$ 0.076 & 1.52 \\
 & 293 & 1.45 $\pm$ 0.14 & 1.93 $\pm$ 0.30 & 0.429 $\pm$ 0.045 & 1.22 \\
 & 288 & 1.50 $\pm$ 0.18 & 2.16 $\pm$ 0.30 & 0.409 $\pm$ 0.044 & 1.32 \\
 & 283 & 1.17 $\pm$ 0.18 & 1.37 $\pm$ 0.35 & 0.461 $\pm$ 0.073 & 1.34 \\
 & 278 & 0.77 $\pm$ 0.17 & 1.00 $\pm$ 0.21 & 0.436 $\pm$ 0.076 & 1.51 \\
 & 273 & 0.40 $\pm$ 0.06 & 0.61 $\pm$ 0.07 & 0.397 $\pm$ 0.046 & 1.25 \\
 & 268 & 0.33 $\pm$ 0.12 & 0.45 $\pm$ 0.09 & 0.424 $\pm$ 0.101 & 1.44 \\
 & 263 & 0.08 $\pm$ 0.02 & 0.11 $\pm$ 0.03 & 0.430 $\pm$ 0.094 & 0.82 \\
 & 258 & 0.08 $\pm$ 0.01 & 0.06 $\pm$ 0.03 & 0.553 $\pm$ 0.132 & 0.75 \\
 & 253 & 0.06 $\pm$ 0.01 & 0.07 $\pm$ 0.03 & 0.458 $\pm$ 0.130 & 0.96 \\
 & 248 & 0.05 $\pm$ 0.01 & 0.06 $\pm$ 0.02 & 0.419 $\pm$ 0.102 & 0.85 \\
 & 243 & 0.04 $\pm$ 0.01 & 0.05 $\pm$ 0.02 & 0.467 $\pm$ 0.133 & 1.09 \\
 & 238 & 0.01 $\pm$ 0.00 & 0.02 $\pm$ 0.02 & 0.323 $\pm$ 0.258 & 0.46 \\
 & 233 & 0.01 $\pm$ 0.01 & 0.02 $\pm$ 0.02 & 0.390 $\pm$ 0.237 & 1.07 \\
\end{longtable}
\endgroup

\begin{figure}[p]
    \centering
    \includegraphics{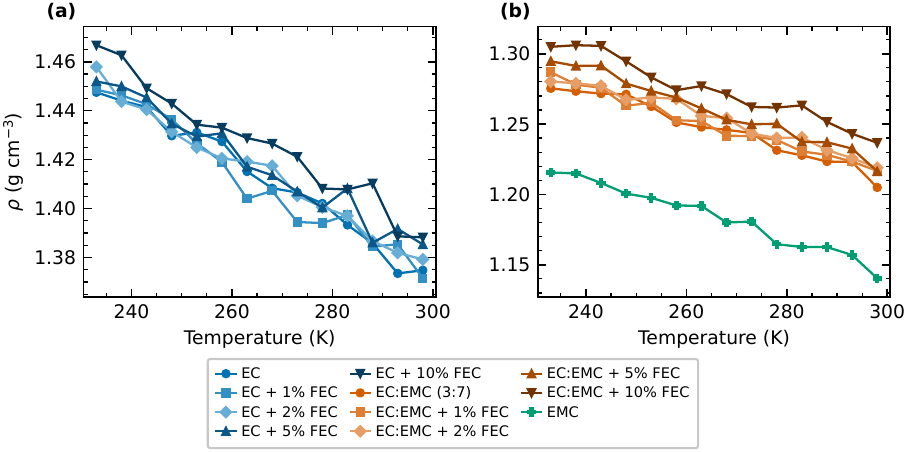}
    \caption{Temperature dependence of the equilibrated mass densities of the simulated 1~M LiPF$_6$ electrolytes: (a) EC-based systems (neat EC and EC with 1-10~mol\% FEC); (b) EC/EMC (3:7, w/w)-based systems and neat EMC. Each point is the NPT-equilibrated density at which the corresponding NVT production run was performed; colors follow the electrolyte color coding used throughout.}
    \label{fig:si_density}
\end{figure}

\end{document}